\documentclass[aps,prd,twocolumn,floatfix,showpacs,superscriptaddress,nofootinbib]{revtex4-1}  
\usepackage{graphicx}  
\usepackage{dcolumn}   
\usepackage{amssymb}   
\usepackage{amsmath}
\usepackage{mathrsfs, dsfont} 
\usepackage[usenames,dvipsnames]{color}
\usepackage{hyperref} 
\usepackage{soul}
\newcommand{\sj}[6]{ \begin{Bmatrix}
     #1 & #2 & #3 \\
     #4 & #5 & #6	 
\end{Bmatrix}}

\newcommand{\tj}[6]{ \begin{pmatrix}
     #1 & #2 & #3 \\
     #4 & #5 & #6	 
\end{pmatrix}}

\mathchardef\mhyphen="2D

\newcommand\undermat[2]{%
    \makebox[0pt][l]{$\smash{\underbrace{\phantom{%
          \begin{matrix}#2\end{matrix}}}_{\text{$#1$}}}$}#2}
\newcommand\overmat[2]{%
    \makebox[0pt][l]{$\smash{\overbrace{\phantom{%
          \begin{matrix}#2\end{matrix}}}^{\text{$#1$}}}$}#2}

\renewcommand{\bm}[1]{\boldsymbol #1}

\newcommand{\apjl}{Astrophys. J. Lett.}
\newcommand{\aap}{Astron. Astrophys.}

\newcommand{\mnras}{Mon. Not. R. Astron. Soc.}

\newcommand{\jcap}{Journal of Cosmology and Astroparticle Physics}
\newcommand{\ssr}{Space Science Reviews}
\newcommand{\aapr}{Astron. and Astrophys. Review}
\newcommand{\solphys}{Solar Physics}
\newcommand{\jqsrt}{J. Quant. Spec. Rad. Trans.}

\newcommand{\supdet}[1]{}

\allowdisplaybreaks 

\makeatletter
\def\l@subsubsection#1#2{}
\makeatother

\begin{document}

\widetext

\title{New probe of magnetic fields in the prereionization epoch. I. Formalism}
\author{Tejaswi Venumadhav}
\affiliation{Theoretical Astrophysics Including Relativity (TAPIR), Caltech M/C 350-17, Pasadena, California 91125, USA}
\affiliation{School of Natural Sciences (SNS), Institute for Advanced Study, Einstein Drive, Princeton, New Jersey 08540, USA}
\author{Antonija Oklop\v ci\'c}
\affiliation{Theoretical Astrophysics Including Relativity (TAPIR), Caltech M/C 350-17, Pasadena, California 91125, USA}
\author{Vera Gluscevic}
\affiliation{School of Natural Sciences (SNS), Institute for Advanced Study, Einstein Drive, Princeton, New Jersey 08540, USA}
\author{Abhilash Mishra}
\affiliation{Theoretical Astrophysics Including Relativity (TAPIR), Caltech M/C 350-17, Pasadena, California 91125, USA}
\author{Christopher M. Hirata}
\affiliation{Center for Cosmology and Astroparticle Physics (CCAPP), The Ohio State University, 191 West Woodruff Lane, Columbus, Ohio 43210, USA}
\date{\today}

\begin{abstract}
We propose a method of measuring extremely weak magnetic fields in the intergalactic medium prior to and during the epoch of cosmic reionization. The method utilizes the Larmor precession of spin-polarized neutral hydrogen in the triplet state of the hyperfine transition. This precession leads to a systematic change in the brightness temperature fluctuations of the 21-cm line from the high-redshift universe, and thus the statistics of these fluctuations encode information about the magnetic field the atoms are immersed in. The method is most suited to probing fields that are coherent on large scales; in this paper, we consider a homogenous magnetic field over the scale of the 21-cm fluctuations. Due to the long lifetime of the triplet state of the 21-cm transition, this technique is naturally sensitive to extremely weak field strengths, of order $10^{-19}$ G at a reference redshift of $\sim 20$ (or $10^{-21}$ G if scaled to the present day). Therefore, this might open up the possibility of probing primordial magnetic fields just prior to reionization. If the magnetic fields are much stronger, it is still possible to use this method to infer their direction, and place a lower limit on their strength. In this paper (Paper I in a series on this effect), we perform detailed calculations of the microphysics behind this effect, and take into account all the processes that affect the hyperfine transition, including radiative decays, collisions, and optical pumping by Lyman-$\alpha$ photons. We conclude with an analytic formula for the brightness temperature of linear-regime fluctuations in the presence of a magnetic field, and discuss its limiting behavior for weak and strong fields.
\end{abstract}

\maketitle


\section{Introduction}
\label{sec:introduction} 

Magnetic fields (MFs) are seen in astrophysical structures on a wide range of observable scales, both in the local universe \cite{2005LNP...664...89W, 2012SSRv..166..215B} and at high redshifts \cite{2008Natur.454..302B}. Typical field strengths in galaxies and galaxy clusters are a few to a few tens of $\mu$G, with coherence lengths of up to hundreds of kpc \cite{Vallee04}. However, properties of the intergalactic MFs on even larger length scales are largely unknown.

The leading paradigm for the origin of large-scale cosmic MFs assumes some kind of amplification and dynamo-based sustaining of weak seed fields \cite{2013A&ARv..21...62D}. These seed fields may originate from mechanisms effective during structure formation, or could be primordial remnants from the early universe (see, for example, Refs.~\cite{2013A&ARv..21...62D,2012SSRv..166...37W, 2013PhRvL.111e1303N,2011ApJ...729...73M,2014JCAP...05..040K}). The search for primordial magnetic fields (PMFs) is an active area of investigation in astrophysics and cosmology, as their observation would open up a new window into the physics of the early universe and possibly provide an entirely unexplored source of information about inflationary and prereheating processes. 

Current upper limits on large-scale MFs come from several different observations, and are on the order of $10^{-9}$ G. They are derived from the limits on Faraday rotation of the cosmic-microwave-background (CMB) polarization \cite{Yamazaki10} and of the radio emission from distant quasars \cite{Blasi99}, measurements of the CMB temperature anisotropies \cite{2013PhLB..726...45P}, limits on CMB spectral distortions \cite{2014JCAP...01..009K}, and various observations of  large-scale structure (LSS) \cite{2013ApJ...770...47K}. 

More recently, observations of TeV sources by the Fermi mission have been interpreted as implying the existence of magnetic fields stronger than $10^{-15}$ G with Mpc scale coherence lengths, in local LSS voids \cite{Neronov10, Tavecchio10, Dolag11}. Plasma instabilities might avoid these bounds by eliminating the expected cascade of lower-energy gamma rays \cite{Broderick12}, but recent calculations indicate these instabilities might saturate, and thus challenge the viability of this argument \cite{Miniati13,Sironi14} (but see also Ref.~\cite{Chang14}). The lower limit may also be reduced if the TeV emission timescale is short, since the arrival of the lower-energy cascade photons is delayed relative to the direct TeV photons \cite{Dermer11,Taylor11}.

Most of these methods are sensitive to the integrated effect of MFs along a line of sight, and thus can be contaminated by low-redshift magnetic fields of astrophysical origin---for instance, those carried by galactic winds (for a notable exception, see Ref.~\cite{2014MNRAS.445L..41T} which probes local fields using statistical correlations in the gamma ray sky). Moreover, these methods optimally detect fields that are stronger than typical expectations for PMFs. Thus a definitive probe of PMFs needs to have the following features: 
\begin{itemize}
  \item The ability to isolate the effects of fields at different redshifts. In particular, sensitivity at high redshifts (prior to, or at the dawn of structure formation). 
  \item Sensitivity to extremely low field strengths. Inflationary, post-inflationary, and structure-formation related mechanisms typically generate seed fields with strengths in the range $10^{-30}$--$10^{-15}$ G \cite{2014JCAP...05..040K, 2013PhRvL.111e1303N}. 
  \item The ability to recover the MF power spectrum, whose features might give insight into the specifics of the process of magnetogenesis. 
\end{itemize}
This is Paper I of a series that proposes a new observational probe of magnetic fields, which has all the desired properties listed above. In this Paper, we lay out the details of the microphysical calculation of the 21-cm signal in presence of the magnetic fields. In Paper II of this series \cite{2016arXiv160406327G}, we present a minimum-variance formalism necessary to estimate the field strength with 21-cm tomography observations, and forecast sensitivity of future experiments to detecting two different MF configurations using this method.

The method discussed here is based on the effect of global MFs on the redshifted 21-cm emission from neutral hydrogen prior to and during the epoch of cosmic reionization (EoR), whose measurement is the goal of a number of low-frequency radio arrays, such as MWA \cite{2011AAS...21813206B}, PAPER \cite{2014ApJ...788..106P}, HERA \cite{2015AAS...22532803D}, LOFAR \cite{2013ASPC..475..377M}, LEDA \cite{2012arXiv1201.1700G}, SKA \cite{2008arXiv0802.1727C}, and others. The 21-cm signal allows insight into very high redshifts (in the approximate range $7 < z < 30$), including early epochs where the intergalactic medium (IGM) was just beginning to be affected by stellar feedback.

This method relies on the availability of internal (spin) degrees of freedom to hydrogen atoms in the triplet state of the ground hyperfine transition. As we show in the body of the paper, an anisotropic radiation field spin-polarizes these levels (also see previous work in Refs. \cite{Varshalovich68, Varshalovich71, Yan06, Yan07, Yan08}). Such anisotropies are naturally present in the early universe due to density fluctuations in the neutral gas. In the presence of a background magnetic field, the Larmor precession of the atoms leads to a characteristic signature in the 21-cm brightness temperature. In particular, a homogenous magnetic field breaks the statistical isotropy of the measured two-point correlation functions of the brightness temperature. This effect is inherently sensitive to extremely weak MFs, smaller than $\sim 10^{-19}$ G at a reference redshift $z \sim 20$.\footnote{Note that a ``frozen'' magnetic field should scale as $\propto(1+z)^2$ due to flux conservation; the comoving field strength, defined by extrapolation to the present day, would be $10^{-21}$ G.} This remarkable sensitivity is due to the long lifetime of the excited state, during which even very slow precession results in a substantial change in the direction of the emitted radiation.

The rest of this paper is organized as follows. We give some background about $21$-cm cosmology and the Hanle effect (which is closely related to the effect considered in this paper) in Secs.~\ref{subsec:21cm} and \ref{subsec:hanle}. We then introduce the effect in a simple, semiclassical manner in Sec.~\ref{sec:estimate}. We lay out the notation and formalism we use in Sec.~\ref{sec:notation}, including our description of spin-polarized atoms in \ref{subsec:atomdm} and the anisotropic radiation field in the vicinity of the $21$-cm transition in \ref{subsec:photondm}. Next, we study the excitation and deexcitation of the atoms by the $21$-cm radiation in Sec.~\ref{sec:dmevolnradiative}. We compute the rates of depolarization by competing nonradiative processes in Sec.~\ref{sec:otherproc}, with \ref{subsec:collisions} and \ref{subsec:opticalpumping} addressing spin-exchange collisions and optical pumping by Lyman-$\alpha$ photons, respectively. We describe the radiative transfer of $21$-cm photons in Sec.~\ref{sec:radtrans}. We put together all these results and calculate the resulting change in the brightness-temperature fluctuations in Sec.~\ref{sec:results}. Finally, we summarize the paper and lay out our conclusions in Sec.~\ref{sec:summary}. Various technical details involved in the computations are collected into the appendices.

\section{Background}
\label{sec:background}

\subsection{21-cm cosmology basics}
\label{subsec:21cm}

The 21-cm line of neutral hydrogen corresponds to the transition between the hyperfine sublevels of its ground state, whose origin is the interaction between the spins of the proton and the electron. This interaction reorganizes the four possible spin states of the electron and proton into singlet and triplet levels, which are separated by an energy gap of $5.9 \times 10^{-6} \ {\rm eV}$, corresponding to radiation with a wavelength of $21.1 \ {\rm cm}$ (or a frequency of $1420 \ {\rm MHz}$), in the rest frame.

In the early stages of the EoR, the universe was still mostly neutral, and fluctuations in the brightness temperature of the 21-cm line were mainly driven by (mostly Gaussian) density fluctuations. This stage lends itself to a very precise statistical description, allowing us to get a good handle on the expected 21-cm signal from the corresponding redshifts \cite{Pritchard12}. 

The first generation of EoR experiments, such as the MWA, PAPER, and LOFAR, aim to achieve a statistical detection of the 21-cm signal from the EoR. Second generation experiments, such as the SKA and future phases of HERA, planned to come online within the next couple of decades, aim to perform detailed tomography of the IGM out to $z\sim 30$. Future 21-cm observations of the high-redshift universe can open up a new frontier in cosmology, with a sample volume far exceeding that probed with current observations. Several authors have suggested that cosmological 21-cm radiation could be used to detect primordial magnetic fields via their dynamical effects on density and gas temperature fluctuations \cite{2006MNRAS.372.1060T, 2009ApJ...692..236S, 2014PhRvD..89j3522S}. The method proposed here \supdet{using radiative transfer }is sensitive to much weaker fields than those investigated by other authors.

The conventional appeal of 21-cm observations is the availability of redshift information (in contrast to other probes of the very early universe, such as the CMB), the access to small-scale modes (Silk damped in the CMB and altered by nonlinear evolution today), and the consequent large number of accessible modes \cite{2004PhRvL..92u1301L}. The effect studied in this paper relies on another aspect of the transition: in the triplet state, the net magnetic moment of the atom (which is dominated by the magnetic moment of the electron), takes on different values depending on the magnetic quantum number. It is through this magnetic moment that the 21-cm emission is sensitive to ambient MFs, as explained in the following sections.

For unpolarized atoms, the detectability of the 21-cm signal hinges on the spin temperature $T_{\rm s}$, which quantifies the
relative number densities of atoms in the two hyperfine levels of the electronic ground state,
\begin{equation}
\frac{n(F=1)}{n(F=0)}= 3 e^{-T_{*}/T_{\rm s}}.
\end{equation}
Here, $F=0$ denotes the lower (spin-antiparallel) hyperfine level, $F=1$ denotes the upper (spin-parallel) level, $3$ is the ratio of statistical weights, and $T_*=\hbar\omega_{\rm hf}/k_B = 68\,$mK is the hyperfine splitting in temperature units.
A signal is detected if the spin temperature of the gas deviates from
the temperature of the background CMB $T_\gamma$ at that redshift; net emission occurs if $T_{\rm s}>T_\gamma$ and absorption if $T_{\rm s}<T_\gamma$. The spin temperature is determined by
three major processes: (1) absorption/emission of 21-cm photons from/to the radio background
at that redshift (primarily the CMB), (2) collisional excitation and deexcitation of hydrogen atoms, and
(3) resonant scattering of Ly$\alpha$ photons from the first stars and galaxies, which can change the spin state via the spin-orbit interaction while the atom is in the excited state.

The fundamental quantity of interest observationally is the {\it brightness temperature}
of the H\,{\sc i} 21-cm line \cite{1997ApJ...475..429M}. In the optically thin approximation, the brightness temperature
{\it fluctuation} relative to the CMB at redshift $z$ and hence observed at the frequency $\omega_{\rm obs} = \omega_{\rm hf}/(1+z)$ is
\begin{equation}
\delta T_{\rm b}\approx 27x_{\rm 1s} (1+\delta) \frac{T_{\rm s}-T_{\gamma}}{T_{\rm s}} \left( \frac{1+z}{10}\right)^{1/2} \frac{(1+z)H(z)}{\partial_\parallel v_\parallel} \,\mathrm{mK}
\end{equation}
(see e.g. Ref.~\cite{Pritchard12}).\footnote{Note that Eq. (7) in Ref.~\cite{Pritchard12} is missing a $-1$ exponent.}
Here $x_{\rm 1s}$ is the hydrogen neutral fraction (essentially all in the ground state), $1+\delta$ is the matter density contrast, $T_{\rm s}$ is the spin temperature, and the line-of-sight velocity gradient $\partial_\parallel v_\parallel$ accounts for deviations from the expansion rate of the homogeneous universe.

In this paper, where we take account of the spin-polarization of atoms, we need to consider the full atomic density matrix, rather than $T_{\rm s}$ alone. We extend the formalism of 21-cm cosmology as needed to derive an equation for $\Delta T_{\rm b}$ valid in this case. Several previous analyses have considered polarized 21-cm radiation from high redshift and its ``scrambling'' by Faraday rotation in passing through the interstellar medium of our own galaxy \cite{2005ApJ...635....1B, 2014PhRvD..89l3002D}; however they did not study polarization of the emitting atoms\footnote{These works focused on polarization produced by re-scattering of 21-cm radiation by electrons in ionized regions. There is no anisotropy of the spins of the hydrogen atoms involved in that mechanism.}, and thus did not need to develop the formalism presented here.

\subsection{Related methods: Hanle effect and ground-state alignment}
\label{subsec:hanle}

The effect considered in this paper is closely related to the Hanle effect \cite{Hanle23}, which refers to the change in the polarization of resonant-scattering radiation in the presence of external MFs. In solar research, techniques based on the Hanle effect are used for measuring weak MFs in solar prominences and the upper solar atmosphere (see e.g. Refs.~\cite{1965ApJ...141.1374H, 1977A&A....59..223S, Bommier78, Stenflo82, Leroy85, DeglInnocenti90, Bommier94}). The methods of this paper use the irreducible tensor approach to the density matrix (see Ref.~\cite{1977PQE.....5...69O} for an overview and Ref.~\cite{Bommier78} for an application to the Hanle effect in solar physics).

The subject of this paper relies on atomic alignment, whose significance in the astrophysical context was first realized in the early days of maser studies. The theory of alignment in astrophysical environments was further developed in the pioneering work of Varshalovich \cite{Varshalovich68,Varshalovich71}. Other significant milestones were the work of Goldreich, Keeley and Kwan \cite{Goldreich73a,Goldreich73b}, who considered the polarization of maser emission due to aligned molecules, and Goldreich and Kylafis \cite{Goldreich81}, who proposed using linear polarization in radio lines as probes of magnetic fields in molecular clouds. 

More recently, Yan and Lazarian \cite{Yan06, Yan07, Yan08} proposed a suite of methods to probe weak MFs in diffuse media using atomic alignment. Since the method discussed in this paper relies on the same atomic physics as these previous studies, we briefly summarize the main idea behind them. Their methods rely on the polarization and intensity of radiation interacting with atoms or ions with fine (or hyperfine) structure in the ground state. When these species are immersed in an anisotropic flux of photons, the orientation of the total atomic angular momentum vector gets a preferred direction since photons carry angular momentum and transfer it via interactions. If aligned atoms are further placed in an external MF, their orientations change due to Larmor precession. As a result, the output radiation's intensity and polarization changes in a manner depending on the direction and strength of the MF. The main advantage of using atomic species with (hyper)fine structure in their ground or metastable states is these states' long lifetimes. Longer lifetimes are associated with longer baselines for Larmor precession, which make the effect sensitive to very weak MFs. These authors recognize the relevance of this effect for studying magnetic fields during the EoR via the 21-cm line of neutral hydrogen \cite{Yan08} and the fine-structure lines of the first metals \cite{Yan12}, but they do not include its calculation in the cosmological context.

The study in this paper distills elements from the physics of all the previous work on astrophysical alignment, and uses features unique to the study of the 21-cm line in the cosmological context in order to synthesize a new method for measuring MFs. In order to align the excited state of the 21-cm transition, our method relies on ``resonant" anisotropies (at frequencies $\nu = 1.42~$GHz) that are sourced by fluctuations of large scale structure (LSS). This is closely related to the mechanism of Refs. \cite{Goldreich73a,Goldreich73b,Goldreich81}, in that it uses anisotropies in optical depth sourced by velocity gradients in order to achieve alignment. The mechanism studied in \cite{Yan06, Yan07, Yan08} aligns the triplet via optical pumping by anisotropies in the incident Lyman-$\alpha$ radiation field, i.e. at frequencies $\nu \approx 2.46 \times 10^{15}~$GHz. 

Our method also differs from these previous methods in the respect that it uses relatively subtle changes in the intensity of the outgoing radiation to detect MFs. References \cite{Yan06, Yan07, Yan08} recognize the change in the net emissivity, and propose using the emissivity ratio of multiple lines to probe MFs. As we show in this paper, it is possible to use {\em solely} the 21cm transition, due to the statistical nature of its measurement in cosmology. The cosmic density field contains perturbation modes with a variety of wave vectors ${\bm k}$, whose amplitudes obey the underlying statistical isotropy of the Universe. The anisotropy in the scattering properties caused by the MF can then be probed using the varying illumination conditions (depending on the direction of $\hat{\bm k}$), rather than the polarization of outgoing radiation.

\section{Illustration and simple estimate of the effect}
\label{sec:estimate}

Consider a hydrogen atom in the ground state of the hyperfine transition, located in the overdense part of a growing Fourier mode at a suitably high redshift. Moreover, let us assume that the $21$-cm line is visible in emission. The brightness temperature fluctuation $\delta T_{\rm b}$ seen by this atom along a particular line of sight (LOS) $\hat{\bm n}$ is largely due to stimulated emission and absorption by a thermal background of excited atoms, and is proportional to the optical depth $\tau$ integrated along that direction,
\begin{equation}
  \delta T_{\rm b}(\hat{\bm n}) \approx \tau(\hat{\bm n}) (T_{\rm s} - T_\gamma) \mbox{,} \label{eq:tbdef} 
\end{equation}
where $T_{\rm s}$ and $T_\gamma$ are the spin- and CMB-temperatures, respectively. 

The optical depth, in turn, depends on the path length over which photons stay within the line:
\begin{equation}
  \tau(\hat{\bm n}) \sim n \int \sigma(\nu) {\rm d}l = n \int \sigma(\nu) \frac{ {\rm d}l }{ {\rm d} \nu } {\rm d}\nu \sim \frac{n \sigma(\nu_0) c \Delta}{ {\rm d} v_{||}/{\rm d} r_{||}(\hat{\bm n})} \mbox{,}
\end{equation}
where $\sigma(\nu)$ is the absorption cross-section at frequency $\nu$, $\nu_0$ is the frequency at line-center, $\Delta$ is the dimensionless Doppler width of the line, $c$ is the speed of light, and ${\rm d} v_{||}/ {\rm d} r_{||}(\hat{\bm n})$ is the velocity gradient along the LOS. The velocity gradient term equals the Hubble rate when the LOS is orthogonal to the wave-vector $\bm k$ of the Fourier mode, but it picks up a contribution from the infall into the growing overdensity when the LOS has a component along $\bm k$. For an arbitrary direction of the LOS, the velocity gradient term equals
\begin{equation}
  \frac{ {\rm d} v_{||}}{ {\rm d} r_{||}}(\hat{\bm n}) = H + \frac{ {\rm d} v_{\rm{infall},||}}{ {\rm d} r_{||}}(\hat{\bm n}) = H \left[1 - (\hat{\bm k} \cdot \hat{\bm n})^2 \delta \right] \mbox{.}
\end{equation}
Hence the optical depth of the medium around the atom has a quadrupole dependence with a fractional size proportional to the overdensity, or an absolute size of $ {\cal O}(\delta \tau)$. This leads to a quadrupole in the incident brightness temperature, oriented such that directions along the wave-vector are hotter.

\begin{figure}[t]
  \centering
  \includegraphics[width=8cm]{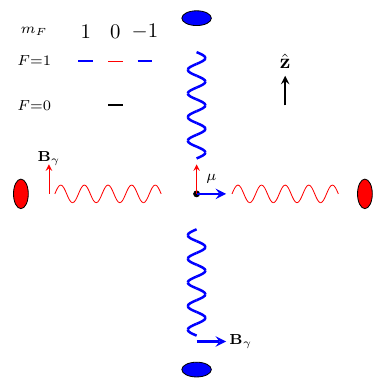}
  \caption{An illustration of how an incident quadrupole spin-polarizes the triplet level of the hyperfine transition. The hydrogen atom (at the center) is surrounded by a quadrupole intensity pattern with hot (blue, thick lines) and cold (red, thin lines) spots. Absorption of $21$-cm photons produces a state with a magnetic moment $\bm \mu$ aligned with the magnetic field $\bm B_\gamma$ of the incident radiation. The incident anisotropy is tranferred to the direction of the magnetic moment. \textit{Inset:} The resulting unequal population of the triplet sublevels. For the orientation of this figure, the levels with magnetic quantum number $m_F = \pm 1$ (thick blue lines) are preferentially populated due to the hot spots.}
  \label{fig:spinpolarization}
\end{figure}

 Atoms that are excited by absorption have magnetic moments that are aligned with the exciting radiation's magnetic field. For anisotropic incident radiation, this leads to a preference for directions orthogonal to that of hot spots in the incident radiation field. Thus an incident quadrupole spin-polarizes the atoms, i.e. unequally populates the states within the hyperfine triplet. Figure \ref{fig:spinpolarization} illustrates this effect.

\begin{figure}[t]
\includegraphics[trim=0cm 0cm 0cm 2cm, clip=true, width=8cm]{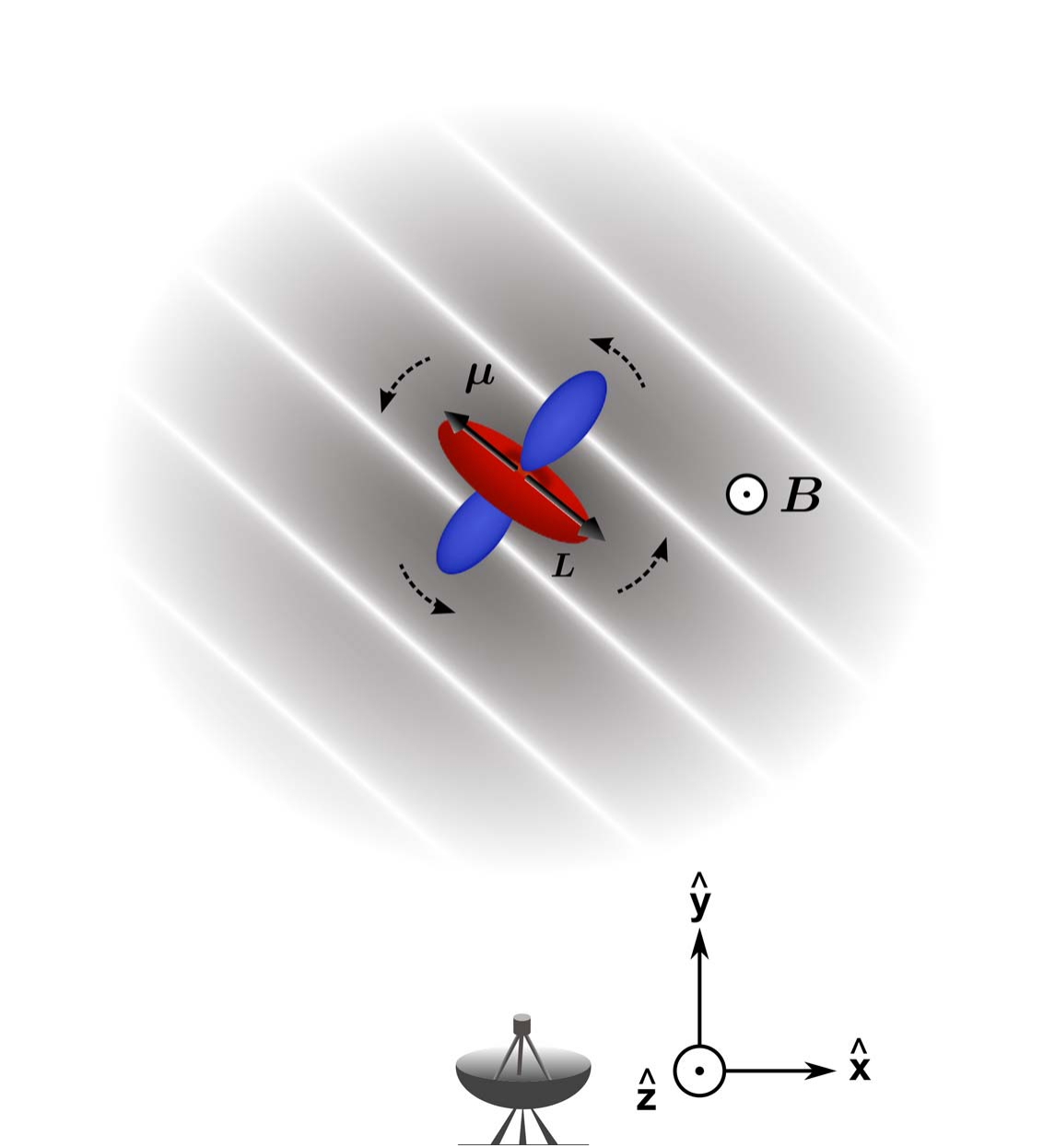}
\caption{A hydrogen atom in a growing plane wave density fluctuation: The atom is excited to the spin-polarized state of Fig.~\ref{fig:spinpolarization}, which produces the quadrupolar radiation pattern shown above when it deexcites. Also shown is one possible orientation of the intermediate magnetic moment $\bm \mu$, and the associated angular momentum $\bm L$. If an external magnetic field $\bm B$ is present, the torque it exerts ($\bm \mu \bm \times \bm B$) causes the moment to precess around it before it deexcites. If the field is coherent on larger scales than the fluctuations in the 21--cm signal (and thus homogenous on the latter's scales), and has a component in the plane of the observer's sky, it systematically changes the brightness temperature for a plane wave as a function of the latter's orientation.}
\label{fig:schematic}
\end{figure}

These excited atoms deexcite to the ground state mainly by stimulated emission or nonradiative processes. The former leads to an output quadrupole pattern with the same orientation as the incident one, but a smaller size of ${\cal O}(\tau\,\delta \tau)$. This is illustrated in Fig.~\ref{fig:schematic}.

The angular structure of the observed brightness temperature fluctuations is dominated by the contribution of the preexisting thermal background of excited atoms, and is $ {\cal O}(\delta \tau)$ in size, as can be seen from Eq.~\eqref{eq:tbdef}. The secondary emission described above is much smaller (by a factor of the optical depth, $\tau$), and does not correspond to a qualitatively different pattern.

The presence of a homogenous background magnetic field breaks isotropy and leads to a unique signature in the angular pattern of this secondary emission. To see this, consider the effect of a magnetic field on the intermediate magnetic moment, which has a finite lifetime $t_{\rm d}$. This lifetime is mainly due to stimulated emission and nonradiative processes such as collisions and optical pumping by Lyman-$\alpha$ photons. Additionally, the moment precesses about the background magnetic field $\bm B$ with the Larmor frequency $\omega_{\rm L}$.

Due to these effects, the moment $\bm \mu$ evolves as
\begin{equation}
  \frac{ {\rm d} }{ {\rm d} t } \bm \mu \approx - \frac{\bm \mu}{ t_{\rm d} } - \omega_{\rm L} \bm \mu \bm \times \widehat{\bm B} \mbox{.}
\end{equation}
In a coordinate system with the background magnetic field along the $z-$axis, the solution is
\begin{equation}
  \bm \mu(t) = e^{-t/t_{\rm d}} 
  \begin{pmatrix}
    \cos{(\omega_{\rm L} t)} & -\sin{(\omega_{\rm L} t)} & 0 \\
    \sin{(\omega_{\rm L} t)} & \cos{(\omega_{\rm L} t)} & 0 \\
    0 & 0 & 1
  \end{pmatrix} \bm \mu_0 \mbox{.}
\end{equation}
Thus the moment precesses through an angle $\theta_{\rm B} \approx \omega_{\rm L} t_{\rm d}$ before the atom deexcites. If the deexcitation occurs only via radiative processes, the lifetime is
\begin{equation}
  t_{\rm d}^{-1} \approx A \frac{k_{\rm B} T_\gamma}{\Delta E_{\rm hf}} \mbox{,}
\end{equation}
where $k_{\rm B}$ is the Boltzmann constant, $\Delta E_{\rm hf}$ is the hyperfine energy gap, and $A$ is the Einstein $A$-coefficient or intrinsic width of the line, which is broadened due to stimulated emission by the background CMB with a temperature $T_\gamma$.

\begin{figure}[t]
  \centering
  \includegraphics[width=8cm]{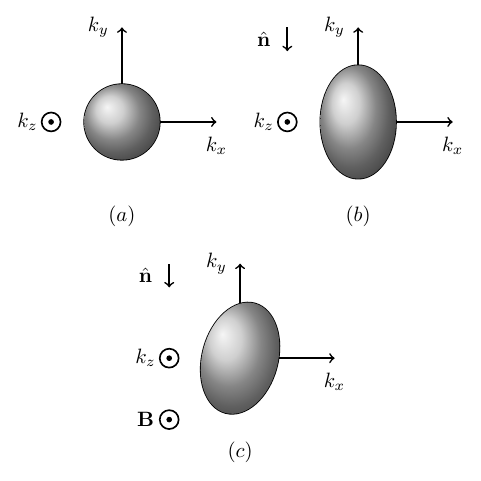}
  \caption{This figure illustrates the effect on the power-spectrum of the brightness temperature fluctuations. The subfigures show contours of constant power in ${\bm k}$-space.
  (a) Fluctuations of the 21-cm emissivity (photons per cm$^3$ per s emitted over all solid angles) in the rest-frame of the emitting atoms.
  (b) Fluctuations as seen by a distant (present-day) observer. Note the elongation in the direction of the line of sight to the observer, $\hat{\bm n}$, due to peculiar velocities. This manifests as a ``compression'' in the real-space correlation function, but as a power enhancement [``stretching'' of the $P({\bm k})$ contours] in Fourier-space.
  (c) Fluctuations with an external magnetic field added. The effect of the precession is to break the symmetry around $\hat{\bm n}$. The size of the effects has been exaggerated in (b) and (c).}
  \label{fig:effectontb}
\end{figure}

We estimate the angle of precession to be
\begin{equation}
  \theta_{\rm B} \approx \omega_{\rm L} t_{\rm d}
  = \frac{\gamma_{\rm e} \Delta E_{\rm hf}}{A k_{\rm B} T_\gamma} B 
  = 1.5 \times \left( \frac{B}{10^{-19} {\rm G}} \right) \left( \frac{1+z}{10} \right)^{-1} \mbox{,}
\end{equation}
where $\gamma_{\rm e}$ is the gyromagnetic ratio of the electron. Figure \ref{fig:schematic} illustrates the precession of the magnetic moment, and that of the quadrupole associated with the secondary emission. If the background magnetic field varies on larger scales than that of the fluctuations in the 21--cm signal, it is effectively homogenous over the length-scale of the modes of interest. In this case, we see from the geometry of the figure (with the magnetic field along the $z-$axis) that the change in a mode's brightness temperature depends on which quadrant of the $x-y$ plane the projection of $\bm k$ lies in. Keeping the line of sight along $\hat{\bm y}$ and assuming the precession angle is small,
\begin{equation}
  \delta T_{\rm b} \vert_{\rm pr} \sim (T_{\rm s} - T_\gamma)\tau\, \delta \tau \left( \theta_{\rm B_z} \hat{\bm k}_x \hat{\bm k}_y - \theta_{\rm B_x} \hat{\bm k}_y \hat{\bm k}_z \right) \mbox{.} \label{eq:estimate}
\end{equation}
The precession-induced correction shown in Eq.~\eqref{eq:estimate} distorts the angular structure of the $21$-cm emission in a manner unlike any of the usually considered effects---it breaks the symmetry around the line of sight. This distinguishes it from corrections like the usual redshift space distortions due to peculiar velocities. Figure \ref{fig:effectontb} illustrates this.

In the rest of the paper we go beyond this simple semiclassical treatment of the spin-polarization, and compute the rates of depolarization by other nonradiative channels.

\section{Notation and Basic Formalism}
\label{sec:notation}

Table \ref{tab:main} lists the symbols used throughout this paper and the physical quantities they represent.

\begin{table*}[t]
\caption{\label{tab:main}Glossary of symbols used in this paper.}
\begin{ruledtabular}
  \begin{tabular}{l p{0.8\textwidth}}
Symbol & Physical quantity\\
\hline
$\rho$  & Density matrix of neutral hydrogen atoms \\
$\rho_{aa}$ & Singlet state submatrix of $\rho$. It is a scalar which corresponds to the occupancy of the singlet state \\
$\rho_{mn}$ & Triplet state submatrix of $\rho$ \\
$\mathscr{P}_{j m}$ & Irreducible components of $\rho_{m n}$\\
$\omega_{\rm hf}$ & Angular frequency of the hyperfine transition \\
$T_*$ & Hyperfine gap expressed in temperature units \\
$A$ & Einstein $A$-coefficient for the hyperfine transition \\
\hline
$k^{\pm}$ & Averaged cross-sections for collisional transitions \\
$\kappa(1 \mhyphen 0)$ & Collisional rate for transition from triplet to singlet state \\
$\kappa(0 \mhyphen 1)$ & Collisional rate for transition from singlet to triplet state \\
$\kappa^{(j)}(1 \mhyphen 0)$ & Collisional depolarization rates for rank-$j$ irreducible components \\
\hline
$n$ & Principal quantum number \\
$l$ & Azimuthal quantum number \\
$m$ & Magnetic quantum number \\
$F$ & Total angular momentum (nuclear $+$ electronic)\\
$m_F$ & Total magnetic quantum number\\
$J_\alpha$ & Flux of Lyman-$\alpha$ photons on the blue side of the line (in $\text{cm}^{-2}\text{s}^{-1}\text{Hz}^{-1}\text{sr}^{-1}$) \\
$T_{\rm c, eff}$ & Effective color temperature in the vicinity of the Lyman-$\alpha$ resonance \\
$\Gamma_{2{\rm p}}$ & Einstein $A$-coefficient for the Lyman-$\alpha$ transition \\
$\gamma_{2{\rm p}}$ & $= \Gamma_{2{\rm p}}/4\pi$, HWHM of the Lyman-$\alpha$ transition \\
$\phi_{\rm AB} (\nu)$ & Interference profiles for the lines $A$ and $B$ \\
$\sigma_{F_I \rightarrow F_J, (j)}(\nu)$ & Cross section for the transition between the rank-$j$ components of multiplets with $F = F_I, F_J$ due to optical pumping by incident Lyman-$\alpha$ photons of frequency $\nu$ \\  
$\tilde{S}_{\alpha}, \tilde{S}_{\alpha, (2)}$ & Correction factors for the detailed frequency dependence of Lyman-$\alpha$ flux, entering the rate equations for $\mathscr{P}_{0 0}$ and $\mathscr{P}_{2 m}$ \\
\hline
$\bm k_\gamma$ & Wave-vector of the radiation \\
$\hat{\bm n}$ & Direction of the radiation's propagation (line-of-sight from the emitter to the observer) \\
$f_{\alpha \beta}(\omega)$ & Phase space density (p.s.d) matrix for the radiation \\ 
$f_{X}(\omega)$ & Parity invariants of the radiation's p.s.d \\
$\mathcal{F}_{j m}(\omega)$ & Irreducible components of the radiation's p.s.d \\
$\phi(\omega)$ & Absorption profile for the hyperfine transition \\
$\mathcal{X}(\omega)$ & Cumulative function for $\phi(\omega)$ \\
$\sigma(\omega)$ & Absorption cross-section for the hyperfine transition \\
$\tau$ & Optical depth of the medium \\
$\delta T_{\rm b}$ & Brightness temperature fluctuation of the 21-cm line relative to the CMB \\ 
$x_{\alpha, (2)}$ & Relative strength of depolarization through optical pumping and radiative channels \\
$x_{ {\rm c}, (2)}$ & Relative strength of depolarization through collisions and radiative channels \\
$x_{\rm B}$ & Relative rates of precession and radiative depolarization \\
\hline
$\delta$ & Local overdensity \\
$\bm v$ & Bulk matter velocity \\
$\bm k$ & Wave-vector of the growing mode of the matter density \\
\hline
$z$ & Redshift \\
$T_{\rm s}$ & Spin temperature \\
$T_\gamma$ & CMB temperature \\
$T_{\rm k}$ & Kinetic temperature \\
$n_{\rm H}$ & Number density of hydrogen atoms \\
$x_{1{\rm s}}$ & Fraction of hydrogen atoms in the $1{\rm s}$ state \\
$H$ & Hubble expansion rate \\
$\bm B$ & External magnetic field in the region of interest \\
\end{tabular}
\end{ruledtabular}
\end{table*}

\subsection{Atomic density matrix}
\label{subsec:atomdm}

We study the level populations of the hydrogen ground state using the density matrix formalism \cite{Blum12}. If we consider an ensemble of atoms consisting of a mixture of states $\vert \psi_\alpha\rangle$ with statistical weights $W_\alpha$, then the density operator is defined as
$\rho = \sum_\alpha W_\alpha \vert \psi_\alpha\rangle \langle \psi_\alpha \vert$.
\supdet{Given the density matrix $\rho$, the expectation value of a general dynamical operator $\mathcal{M}$ is
\begin{equation}
  \langle \mathcal{M} \rangle = \text{Tr} \left[ \rho \mathcal{M} \right] \mbox{.}
\end{equation}}
In order to express the density operator in matrix form, we choose a set of basis states $\vert \phi_I\rangle$;
the matrix elements of $\rho$ are then given by
\begin{equation}
\rho_{I J} = \langle \vert \phi_J\rangle \langle \phi_I \vert \rangle = \sum_\alpha W_\alpha \langle \phi_I \vert \psi_\alpha \rangle \langle \psi_\alpha \vert \phi_J \rangle \mbox{.} \label{eq:rhomn}
\end{equation}
The interaction between the electronic and the nuclear spin splits the ground state of the hydrogen atom into a superposition of two hyperfine levels, a singlet with quantum numbers $(F=0, m_F=0)$, and a triplet with $(F =1, m_F=0, \pm 1)$. As long as we consider the subset of neutral hydrogen atoms in the $1{\rm s}$ electronic state, these states form a complete basis. In the ket notation, these states are represented by $\vert F m_F \rangle$.

We will henceforth adopt the convention that indices of the kind $I,J,\dots$, when used as subscripts for the density matrix $\rho$ or as state labels, run over all four of the hyperfine states of the $1{\rm s}$ type. They are purely abstract indices. Depending on the context, their instantiations are either the lower-case roman letters $a,b,c$, and $d$ or the numbers $1,0$, and $-1$. Table \ref{tab:symbols} maps the various indices to states. Note that numerical subscripts, referred to by $m,n,\dots$ in the text, run over only the triplet states. They equal the magnetic quantum numbers of the respective states. Thus summations over these numeric indices represent ones over only the triplet states.

Within the basis of the two hyperfine levels, the density matrix is of the form
\begin{align}
  \rho = \rho_{IJ} =& 
  \left(
  \begin{array}{c c}
    \overmat{1 \times 1}{\rho_{a a}} & \rho_{a m} \\
    \rho_{m a} & \undermat{3 \times 3}{\rho_{m n}}
  \end{array}
  \right) \mbox{.} \label{eq:dmatrix} \\
  \notag
\end{align}
This density matrix consists of four submatrices. The upper diagonal submatrix has only one element ($\rho_{a a}$) that describes the probability of finding an atom in the singlet state. The lower diagonal submatrix describes the triplet state. Its diagonal elements represent the probabilities of finding atoms with $F=1$ in the states with the corresponding quantum number $m_F$. The off-diagonal elements describe coherences between states of different $m_F$. The remaining two submatrices, with elements in the first row or column describe the interference between $F=0$ and $F=1$ levels. The time evolution of these terms is proportional to $\exp{(i \omega_{\rm hf} t)}$, where $\omega_{\rm hf} = 2\pi \times 1420$~MHz is the angular frequency corresponding to the hyperfine gap. These terms rapidly oscillate on macroscopic timescales with average values of zero, thus we do not need to follow them in the calculation. 

The processes we are interested in only redistribute atoms between the levels, hence the trace of the density matrix is preserved by them. The trace can be taken to be unity as long as we are interested in the population of atoms in the ground electronic state i.e. $\rho_{a a} + {\rm Tr} (\rho_{m n}) = 1$. 

The $4 \times 4$ Hermitian matrix $\rho$ is described by sixteen real numbers. Removing the six real degrees of freedom constituting the submatrix $\rho_{am}$, and the singlet submatrix $\rho_{aa}$, leaves nine real numbers describing the triplet state submatrix $\rho_{mn}$.

In order to take advantage of the symmetries of the problem, it is convenient to express the density matrix in terms of irreducible tensor operators.  We construct irreducible components of ranks $j=\{0, 1, 2\}$ from the elements of the triplet submatrix, in the manner of Ref.~\cite{Lifshitz71}: \footnote{Note that the definition in Ref.~\cite{Lifshitz71} differs from ours by a factor of $i^j$, due to their usage of a different convention for spherical tensors.}
\begin{eqnarray}
\mathscr{P}_{j m} &=& \sqrt{3(2j+1)}\sum_{m_1, m_2 }(-1)^{1-m_2}\tj{1}{j}{1}{-m_2}{m}{m_1} 
\nonumber \\ && \times \rho_{m_1m_2} \mbox{,}
\label{eq:polmom}
\end{eqnarray}
where the expression in large parentheses is the Wigner 3-j symbol. The indices $j$ and $m$ indicate that the irreducible component $\mathscr{P}_{j m}$ transforms in the same way as the corresponding spherical harmonic $Y_{j m}$ does under a rotation of the axes -- only components with the same rank $j$ mix. The Hermiticity of the density matrix leads to the characteristic behavior of these components under complex conjugation:
\begin{equation}
\mathscr{P}_{j -m} = (-1)^m \mathscr{P}_{j m}^\ast \mbox{.}
\end{equation}
The components of rank zero, one and two are described by one, three and five real numbers respectively. As expected, both descriptions of the triplet state density submatrix have the same total number of real degrees of freedom.

\begin{table}[t]
\caption{\label{tab:symbols}Notation for hyperfine states.}
\begin{ruledtabular}
\begin{tabular}{l c r}
$\vert F m_F \rangle$ & Roman & Numeric\\
\hline
$\vert 0 \hspace{5pt} 0 \rangle$  & a & - \\
$\vert 1 -1 \rangle$ & b & -1 \\
$\vert 1 \hspace{5pt} 0 \rangle$ & c & 0 \\
$\vert 1 \hspace{5pt} 1 \rangle$ & d & 1
\end{tabular}
\end{ruledtabular}
\end{table}

We recover the density matrix in the standard basis from the irreducible components using the following relation:
\begin{equation}
\rho_{m_1m_2} = \sum_{j m}\sqrt{\frac{2j+1}{3}} (-1)^{1-m_2}\tj{1}{j}{1}{-m_2}{m}{m_1} \mathscr{P}_{j m} \mbox{.}
\label{eq:polmom2}
\end{equation}
\supdet{The explicit forms of the irreducible components are as follows:
\begin{subequations}
  \label{eq:polmoments}
  \begin{align}
    \mathscr{P}_{0 0} =& ~ \rho_{1 1} + \rho_{0 0} + \rho_{-1 -1} = {\rm Tr}(\rho_{m n}) \mbox{,} \\
    \mathscr{P}_{1 1} =& ~ - \sqrt{\frac{3}{2}} ( \rho_{0 1} + \rho_{-1 0} ) \mbox{,} \notag \\ 
    \mathscr{P}_{1 0} =& ~ \sqrt{\frac{3}{2}} ( \rho_{1 1} - \rho_{-1 -1} ) \mbox{,} \\  
    \mathscr{P}_{1 -1} =& ~ \sqrt{\frac{3}{2}} ( \rho_{1 0} + \rho_{0 -1} ) \mbox{,} \notag \\
    \mathscr{P}_{2 2} =& ~ \sqrt{3} \rho_{-1 1} \mbox{,} \notag \\
    \mathscr{P}_{2 1} =& ~ - \sqrt{\frac{3}{2}} ( \rho_{0 1} - \rho_{-1 0} ) \mbox{,} \notag \\ 
    \mathscr{P}_{2 0} =& ~ \frac{1}{\sqrt{2}} ( \rho_{1 1} - 2 \rho_{0 0} + \rho_{-1 -1} ) \mbox{,} \\ 
    \mathscr{P}_{2-1} =& ~ \sqrt{\frac{3}{2}} ( \rho_{1 0} - \rho_{0 -1} ) \mbox{, ~~and} \notag \\
    \mathscr{P}_{2 -2} =& ~ \sqrt{3} \rho_{1 -1}  \mbox{.} \notag
  \end{align}
\end{subequations}}
The operator of rank zero is a scalar representing the net probability of finding an atom in the triplet, or $F=1$, state.: $\mathscr{P}_{00}={\rm Tr}(\rho_{m n})$. The operator of rank one is a vector with three components, and is often called the {\em orientation vector}. It is proportional to the internal angular momentum of the ensemble. The operator of rank two is the so-called {\em alignment tensor}, which has five components that are quadratic in angular momentum and has the symmetry of an electric quadrupole.

In many applications, excitations between the singlet and the triplet are isotropic. In such cases, only the operator of rank zero, or the net excited-state occupancy, is relevant. The scenario of interest in this paper involves anisotropic excitations, thus we need to use operators of higher rank to describe the spin state of the atoms, which are said to be {\em spin-polarized}.

For a system in equilibrium with a heat bath with temperature $T$, the elements of the density matrix take the form
\begin{equation}
\rho_{IJ}^{\rm th} = \frac{e^{-\beta E_I}}{Z}\delta_{IJ} \mbox{,} \label{eq:equilibrium}
\end{equation}
where $\beta = (k_BT)^{-1}$, and 
$Z = \sum_I e^{-\beta E_I}$
is the partition function of the ensemble.

Given a general density matrix $\rho_{IJ}$, the spin temperature $T_s$ is defined using this equilibrium formula:
\begin{equation}
  \frac{\mathscr{P}_{00}}{1 - \mathscr{P}_{00}} = \frac{\rho_{1 1} + \rho_{0 0} + \rho_{-1 -1}}{\rho_{aa}} = 3 e^{-(\hbar \omega_{\rm hf}/k_{\rm B} T_{\rm s})} \mbox{.}
\end{equation}
In the regimes of interest, the spin temperature is much larger than the temperature associated with the gap, which is $T_* = \hbar \omega_{\rm hf}/k_{\rm B} = 68.2$ mK. In this limit, the occupancy of the excited state is
\begin{equation}
  \mathscr{P}_{00} \approx \frac{3}{4} - \frac{3 T_*}{16 T_{\rm s}} \mbox{.} \label{eq:spintemperature}
\end{equation}

\subsection{Phase-space density matrix for radiation}
\label{subsec:photondm}

In this section and subsequent sections, we use the Coulomb gauge to describe the electromagnetic field, in which $\nabla\cdot{\bm A}=0$ and there is no scalar potential associated with the radiation. The phase space distribution of the radiation and its multipole decomposition -- including the description of the linear polarization in ``E'' and ``B'' modes -- follows the development in the CMB literature \cite{1997PhRvD..55.1830Z, 1997PhRvD..55.7368K}.

As long as we can approximate the electromagnetic field as Gaussian, we can describe its general state by a density matrix, in the same manner as the spin-states of the hydrogen atoms in Sec.~\ref{subsec:atomdm}. We explicitly realize this by expanding the vector potential in the plane-wave basis:
\begin{equation}
  \bm{A}({\bm r}) = \sum_{\bm k_\gamma, \alpha} \left[ a_\alpha (\bm{k_\gamma}) \bm{A}_{\bm k_\gamma, \alpha}({\bm r}) + a^\dagger_\alpha (\bm k_\gamma) \bm{A}^*_{\bm k_\gamma,\alpha}({\bm r}) \right] \mbox{,} \label{eq:aexpand}
\end{equation}
with mode functions given by
\begin{equation}
  \bm{A}_{\bm k_\gamma, \alpha}({\bm r}) = \left( \frac{2\pi \hbar c^2}{\omega} \right)^{1/2} \bm{e}_{(\alpha)}(\hat{\bm k}_\gamma) e^{i \bm k_\gamma \cdot \bm r} \ \mbox{,} \ \alpha = \pm \mbox{,}
\end{equation}
where $\bm k_\gamma$ is the radiation's wave-vector. The subscript on the wave-vector distinguishes it from that of the density fluctuations. The summation over $\bm k_\gamma$ is shorthand for the integral $\int d^3 \bm k_\gamma/(2\pi)^3$, and the angular frequency is given by $\omega = c k_\gamma$. The symbol $\bm{e}_{(\pm)}(\hat{\bm k}_\gamma)$ represents right- and left-circularly polarizated states, respectively. In terms of the unit vectors $\hat{\bm\theta}$ (north-south polarization) and $\hat{\bm\phi}$ (east-west polarization):
\begin{equation}
  \bm{e}_{(\pm 1)}(\hat{\bm k}_\gamma) = \mp \frac{1}{\sqrt{2}} ( \hat{\bm \theta} \pm i \hat{\bm \phi} ) \rvert_{(\theta,\phi) = (\theta_{k_\gamma}, \phi_{k_\gamma})} \ \mbox{.}
\end{equation}
The expansion coefficients in Eq.~\eqref{eq:aexpand} are annihilation and creation operators for photons with momentum $\hbar \bm k_\gamma$, with canonical commutation relations.

We define the density matrix for radiation in a manner almost exactly paralleling that of Eq.~\eqref{eq:rhomn}, which defined it for the atoms:
\begin{equation}
  \langle a^\dagger_\alpha(\bm{k}_\gamma) a_\beta(\bm{k}^\prime_\gamma) \rangle = (2\pi)^3 \delta(\bm{k}_\gamma - \bm{k}^\prime_\gamma) f_{\beta \alpha}(\omega, \hat{\bm n} = \hat{\bm k}_\gamma) \mbox{,} \label{eq:falphabeta}
\end{equation}
where $\hat{\bm n}$ denotes the direction of propagation. In the general polarized case, the phase-space density matrix for the photons, $f_{\alpha \beta}(\omega, \hat{\bm n})$, is of the form:
\begin{align}
  f_{\alpha \beta} = \begin{pmatrix} f_{++} & f_{+-} \\ f_{-+} & f_{--} \end{pmatrix} 
    = & \, \begin{pmatrix} f_{\rm I} + f_{\rm V} & - f_{\rm Q} + i f_{\rm U} \\ - f_{\rm Q} - i f_{\rm U} & f_{\rm I} - f_{\rm V} \end{pmatrix} \mbox{.}
  \label{eq:photondm}
\end{align}
The decomposition of the phase-space density matrix in Eq.~\eqref{eq:photondm} connects it to the Stokes parameters:
\begin{equation}
  X(\omega, \hat{\bm n}) = \frac{\hbar}{c^2} \frac{\omega^3}{4\pi^3} f_{X}(\omega, \hat{\bm n}), \qquad X \in\{{\rm I,Q,U,V}\},
  \label{eq:stokes}
\end{equation}
where the quantities are defined per unit angular frequency $\omega$.

The elements of the phase-space density matrix transform differently under a rotation of the axes. The diagonal elements are scalars, while the off-diagonal elements are quantities with spin weights of $\pm 2$ \cite{HuWhite97}. Hence, they are decomposed into moments as follows:
\begin{equation}
  f_{\alpha \beta}(\omega,\hat{\bm n}) = \sum_{j,m} \sqrt{\frac{4\pi}{2j+1}} (f_{\alpha \beta})_{j m}(\omega) \left[{_{\alpha - \beta}}Y_{j m}(\hat{\bm n})\right]^* \mbox{.} \label{eq:momentexpansion}
\end{equation}
The quantity ${_s}Y_{j m}(\hat{\bm n})$ is the spin-weighted spherical harmonic with spin-weight $s$. The convention of Eq.~\eqref{eq:momentexpansion} is slightly different from that in the standard cosmology literature. Appendix \ref{sec:conventions} expands on the difference and the reason for adopting the current convention.

Inversion of the coordinate axes (a parity transformation) transforms quantities with spin weights of $\pm 2$ into each other. We further split the moments into parity invariants as follows:
\begin{subequations}
\label{eq:photonparitymoments}
\begin{align}
  (f_{++/--})_{j m} = & \, f_{{\rm I}, j m} \pm f_{{\rm V}, j m} \ \mbox{,} \\ 
  (f_{+-/-+})_{j m} = & \, - f_{{\rm E}, j m} \pm i f_{{\rm B}, j m} \ \mbox{.}
  \end{align}
\end{subequations}
The quantities $f_{{\rm I}, j m}$ and $f_{{\rm V}, j m}$ are moments of the intensity and circular polarization respectively. A parity transformation multiplies the quantities $f_{{\rm E}, j m}$ and $f_{{\rm B}, j m}$ by factors of $(-1)^j$ and $(-1)^{j+1}$ respectively.

In this section, we used the plane wave basis to define the phase-space density matrix and its moments. The interaction term between the atoms and radiation is particularly simple when the EM field is expressed in the spherical wave basis \cite{Bommier78}. Hence we use this basis in the calculation of the evolution of the atomic density matrix due to interaction with radiation.

Appendix \ref{sec:sphwave} expands on the details of the spherical wave basis, and the steps involved in moving back and forth between it and the plane wave basis.

\section{Interaction Between Hydrogen Atoms and 21-cm Radiation}
\label{sec:dmevolnradiative}

In this section, we work out the effect of radiative transitions to and from spin-polarized states of the hydrogen atom. We generalize the usual treatment of absorption, and spontaneous and stimulated emission from the level occupancies to the full density matrix $\rho$. Our description of the atom-radiation interaction Hamiltonian is similar, in principle if not in detail, to Sections 14.1 and 15.4 of Mandel and Wolf \cite{Wolf95}.


The dominant interaction is via a magnetic dipole, and involves the emission or absorption of $j=1$ photons of the magnetic type. The transition matrix element between an initial state $I$, and a final state $J$, is \cite{Lifshitz71}
\begin{equation}
  V_{J I, m}(\omega) = - i \sqrt{\frac{2}{3\pi}} \left(\frac{\hbar \omega^3}{c^3}\right)^{1/2} \Bigl[ -e \bigl\{ Q_{1, m}^{(M)} \bigr\}_{J I} \Bigr] \mbox{,} \label{eq:vabsorption}
\end{equation}
where the angular frequency $\omega$ and the magnetic quantum number $m$ describe the photon absorbed in the process, and $\bigl\{ Q_{1 m}^{(M)} \bigr\}_{J I}$ is a spherical component of the magnetic dipole moment $\bm Q^{(M)}_{J I}$.  The magnetic dipole moment is related to the electron's spin-angular momentum by the gyromagnetic ratio i.e. $- e \, \bm Q^{(M)} = -(g_{\rm e} \mu_{\rm B}/\hbar) \bm S_{\rm e}$, where $g_{\rm e}$ is the Land\'{e} g-factor for the electron spin and $\mu_{\rm B}$ is the Bohr magneton. 

The initial state is the singlet state $a$, the final state lies within the triplet, and the index $m$ is fixed by angular momentum conservation. In order to make this clearer, we substitute these states in Eq.~\eqref{eq:vabsorption} and rewrite it in the form
\begin{equation}
  V_{m_F a, m}(\omega_{\rm hf}) = i \hbar \sqrt{ \frac{A}{2\pi} } \delta_{m m_F} \mbox{,} \label{eq:vintermsofa}
  \end{equation}
where $A= 2.86 \times 10^{-15}\,$s$^{-1}$ is the Einstein coefficient for the hyperfine transition.


Given the transition matrix element, the atom-radiation interaction Hamiltonian is\footnote{Compare Eq.(15.4-3) of Ref.~\cite{Wolf95}. Their interaction Hamiltonian is for a single plane wave mode of the radiation field, and is written in the interaction rather than the Heisenberg picture.}
\begin{equation}
  H_{\rm{hf},\gamma} = \sum_{m_F m} \int d\omega \ V_{m_F a, m}(\omega) \vert 1 m_F \rangle \langle 0 0 \vert a_{1 m}^{(M)}(\omega) + \rm{h.c.} \label{eq:M1Hamilt}
\end{equation}
Here ``h.c." stands for the Hermitian conjugate. The quantity $a_{1 m}^{(M)}(\omega)$ is an annihilation operator for a photon of the magnetic type, expanded upon in Appendix \ref{sec:sphwave}.

From here onwards, we use a dot over a quantity to represent its time evolution. Equation \eqref{eq:rhomn} enables us to write down the evolution of the triplet state submatrix $\rho_{m n}$ due to the interaction with the EM field. The underlying operator commutes with the matter Hamiltonian, so its evolution is solely due to the interaction $H_{\rm{hf},\gamma}$, specifically:
\begin{align}
  ~~ & \!\!\! 
  \dot{\rho}_{m_1 m_2} \rvert_{\gamma} \notag \\
  &= \frac{i}{\hbar} \left\langle \left[ H_{\rm{hf},\gamma}, \vert 1 m_2 \rangle \langle 1 m_1 \vert \right] \right\rangle \notag \\
  &= \frac{i}{\hbar} \sum_{m} \int d\omega V_{m_2 a, m}^*(\omega) \left\langle \vert 0 0 \rangle \langle 1 m_1 \vert a_{1 m}^{(M)}{^\dagger}(\omega) \right\rangle + \rm{c.c.s.} \label{eq:dmrate}
\end{align}
Here ``c.c.s.'' stands for complex conjugation with a swap (i.e. swap $m_1 \leftrightarrow m_2$). 

The three-point functions of the atom and the radiation field represent transitions between the singlet and the triplet levels. Appendix \ref{sec:3ptfn} derives expressions for such three-point functions. Plugging in Eq.~\eqref{eq:3ptfna} gives the evolution equation
\begin{align}
  \dot{\rho}_{m_1 m_2} \rvert_{\gamma} & = - \frac{\pi}{\hbar^2} \!\! \sum_{m, m^\prime, m_3} \!\!
  V^*_{m_2 a, m} V_{m_3 a, m^\prime} \notag \\
  & ~~~\times \Bigl[ \rho_{m_1 m_3} \Bigl\{ \delta_{m m^\prime} + f_{m^\prime, m}^{(M 1)(M 1)} \Bigr\} \notag \\
  & ~~~~~~~
  - \delta_{m_3 m_1} \rho_{a a} \, f_{m^\prime, m}^{(M 1)(M 1)} 
  \Bigr] + \rm{c.c.s.} \label{eq:rho1dotunsimpl}
\end{align}
This uses the notation for the radiation's phase-space density matrix in the spherical basis, defined in Eq.~\eqref{eq:rhojmjm} of Appendix \ref{sec:sphwave}. The transition matrix elements and phase-space density moments are evaluated at $\omega_{\rm hf}$, the angular frequency of the hyperfine transition. However, the frequency in the bulk-rest frame corresponding to $\omega_{\rm hf}$ in the interacting atoms' frame is distributed over a broadened profile due to the thermal motions of the atoms.

In this calculation, we assume that the atom density matrix is independent of the velocity. The practical consequence of this assumption is that Eq.~\eqref{eq:rho1dotunsimpl} can be used as is, with the radiation's phase space density averaged over a Doppler-broadened profile centered around $\omega_{\rm hf}$. The consequences of relaxing this assumption have been explored in a different context before \cite{HirataSigurdson07}. In subsequent equations, a bar over quantities is used to indicate averages over the line profile.

In order to simplify the evolution given by Eq.~\eqref{eq:rho1dotunsimpl}, it is convenient to divide the terms into spontaneous and stimulated emission, and photo-absorption contributions.

Spontaneous emission is described by the terms in Eq.~\eqref{eq:rho1dotunsimpl} connecting the excited state density submatrix $\rho_{m n}$ to itself. We write these terms in terms of the irreducible components $\mathscr{P}_{j m}$ using Eqs.~\eqref{eq:polmom} and \eqref{eq:polmom2}:
\begin{align}
  \dot{\mathscr{P}}_{j m} \rvert_{\rm sp.em} =& - A \mathscr{P}_{j m} \mbox{.} \label{eq:spontem} 
\end{align}
Absorption is described by the terms in Eq.~\eqref{eq:rho1dotunsimpl} connecting the excited state density submatrix $\rho_{m n}$ to the ground state occupancy $\rho_{a a}$. Using Eq.~\eqref{eq:vintermsofa}, we write this contribution as
\begin{equation}
  \dot{\rho}_{m_1 m_2} \rvert_{\rm{ab}} = A \ \rho_{a a} \, \overline{ f_{m_1, m_2}^{(M 1)(M 1)} } \mbox{.}
\end{equation}
We can define irreducible components, $\mathcal{F}_{j m}$, of the M1--M1 block of the photon phase-space density matrix in the same manner as those of the triplet state density submatrix [see Eq.~\eqref{eq:fjm}]. The photo absorption contribution retains its form when expressed in terms of the irreducible components:
\begin{equation}
  \dot{\mathscr{P}}_{j m} \rvert_{\rm{ab}} = A \ \rho_{a a} \ \overline{\mathcal{F}_{j m}} = A \left( 1 - \mathscr{P}_{0 0} \right) \overline{\mathcal{F}_{j m}} \mbox{.} \label{eq:photoabs}
\end{equation}
Stimulated emission is described by the terms in Eq.~\eqref{eq:rho1dotunsimpl} connecting the excited state density submatrix $\rho_{m n}$ to itself, via the photon phase-space density moments $f_{m, n}^{(M 1)(M 1)}$. Using Eq.~\eqref{eq:vintermsofa}, this contribution is
\begin{equation}
  \dot{\rho}_{m_1 m_2} \rvert_{\rm st.em} = - \frac{A}{2} \sum_{m_3} \
  \rho_{m_1 m_3} f_{m_3, m_2}^{(M 1)(M 1)} + \rm{c.c.s.}
\end{equation}
Using Eqs.~\eqref{eq:polmom}, \eqref{eq:polmom2} and \eqref{eq:fjminv}, we rewrite this in terms of the irreducible components $\mathscr{P}_{j m}$ and $\mathcal{F}_{j m}$:
\begin{align}
  ~~ & \!\!\!
  \dot{\mathscr{P}}_{j m} \rvert_{\rm st.em} \notag \\
  & = - \frac{A}{2} \!\! \sum_{m_1 m_2 m_3} \sum_{j^\prime m^\prime j^{\prime\prime} m^{\prime\prime}} \!\! \sqrt{\frac{(2j+1)(2j^\prime + 1)(2j^{\prime\prime}+1)}{3}}  \notag \\
  & ~~~
  \times (-1)^{1-m_3} \Bigl[ \tj{1}{j}{1}{-m_2}{m}{m_1} \tj{1}{j^\prime}{1}{-m_3}{m^\prime}{m_1} \notag \\
  & ~~~
  \times \tj{1}{j^{\prime\prime}}{1}{-m_2}{m^{\prime\prime}}{m_3} + (j^{\prime} m^{\prime} \leftrightarrow j^{\prime\prime} m^{\prime\prime} ) \Bigr] \mathscr{P}_{j^\prime m^\prime} \overline{ \mathcal{F}_{j^{\prime\prime} m^{\prime\prime}} } \mbox{.}
\end{align}
The summations over angular indices for products of three 3-j symbols, when evaluated, yield the product of a Wigner 6-j symbol along with a 3-j symbol \cite{Varshalovich88}. Thus the evolution of the irreducible components $\mathscr{P}_{j m}$ due to stimulated emission is
\begin{align}
  \dot{\mathscr{P}}_{j m} \rvert_{\rm st.em}
  =& - A \sum_{j^\prime, j^{\prime\prime}} \sqrt{\frac{(2j^\prime+1)(2j^{\prime\prime}+1)}{3}} \sj{j^{\prime\prime}}{j^{\prime}}{j}{1}{1}{1} \notag \\
  &
  \times \Bigl[ \frac{ (-1)^j + (-1)^{j^{\prime\prime} - j^{\prime}}}{2} \Bigr] (\mathscr{P}_{j^{\prime}} \otimes \overline{\mathcal{F}_{j^{\prime\prime}} })_{j m} \mbox{.} \label{eq:stimem}
\end{align}
The expression enclosed in curly braces is the 6-j symbol, and the notation $(\mathscr{P}_{j_1} \otimes \mathcal{F}_{j_2})_{j m}$ denotes the sum of products of the irreducible quantities $\mathscr{P}_{j_1 m_1}$ and $\mathcal{F}_{j_2 m_2}$, weighted with appropriate 3-j symbols, to yield a quantity which transforms in the $(j m)$ representation.

In the absence of a density fluctuation, the excited states are isotropically occupied. Thus only the irreducible moment  $\mathscr{P}_{0 0}$ has a zeroth-order contribution. The radiation field is unpolarized in this case, so only the intensity monopole has a zeroth-order contribution. Thus the only relevant radiation moment in the unperturbed case is $\mathcal{F}_{0 0}$.

As discussed in Sec.~\ref{sec:estimate}, a growing density fluctuation leads to an incident quadrupole on the atoms. Hence the extra radiation moment exciting the atoms is of the $\mathcal{F}_{2 m}$ type. The spin-polarization due to this quadrupole is described by the alignment tensor $\mathscr{P}_{2 m}$. The orientation tensor $\mathscr{P}_{1 m}$ can be neglected to the first order in the fluctuations. (The CMB dipole in the baryon rest frame is first-order in perturbation theory, and thus in principle should be considered -- however it has the wrong parity to contribute to $\mathscr{P}_{1m}$.)

When we sum up the contributions of absorption and emission from Eqs.~\eqref{eq:spontem}, \eqref{eq:photoabs} and \eqref{eq:stimem}, we get the net rate of change of the atom density matrix due to radiative processes. Using explicit expressions for the irreducible components $\mathcal{F}_{j m}$ of the phase-space density matrix from Eq.~\eqref{eq:fjm}, we find that
\begin{subequations}
  \label{eq:atomdmevol}
  \begin{align}
    \dot{\mathscr{P}}_{0 0} \rvert_{\gamma}
    =& - A \left[ \mathscr{P}_{0 0} - \left( 3 - 4 \mathscr{P}_{0 0} \right) \overline{f_{\text{I},0 0}} \right] \mbox{~~and} \label{eq:p00evolrad} \\
    \dot{\mathscr{P}}_{2 m} \rvert_{\gamma}
    =& - A \Bigl[ \left( 1 + \overline{f_{\text{I},0 0}} \right) \mathscr{P}_{2 m} - \frac{3 - 4 \mathscr{P}_{0 0}}{5\sqrt{2}} \notag \\
    & \times \left( \overline{f_{\text{I}, 2 m}} + \sqrt{6} \, \overline{f_{E, 2 m}} \right) \Bigr] \mbox{.} \label{eq:p2mevolrad}
  \end{align}
\end{subequations}

\section{Other Processes Affecting the Atomic Density Matrix}
\label{sec:otherproc}

The level populations or spin-polarization of the hydrogen ground state can be altered by mechanisms other than emission/absorption of the 21-cm photons. The ones relevant to the subject of this paper are background magnetic fields, hydrogen-hydrogen collisions, optical pumping by Lyman-$\alpha$ photons. Of these, the effect of the magnetic fields is simplest to evaluate. 

The transition rates for the isotropically occupied cases due to the other processes have been calculated previously \cite{Hirata06, Zygelman05}. In this section, we generalize these results to the case of spin-polarized hydrogen atoms -- in particular we calculate the rates of depolarization due to collisions and optical pumping, which are important for determining the lifetime of the excited state of Sec.~\ref{sec:estimate}.

\subsection{Background magnetic field}
\label{subsec:magfield}

We choose the coordinate system such that the $z$-axis is oriented along the external magnetic field; in this case, the $F=1$ states are split and the effect of the external field is to contribute a correction $\Delta E = g_e\mu_{\rm B}Bm_F/2$ to the energy of each state. The elements of the density matrix then vary as $\dot\rho_{m_1m_2} = i (g_{\rm e} \mu_{\rm B}B/2\hbar) (m_2-m_1)\rho_{m_1m_2}$, and the irreducible components $\mathscr{P}_{j m}$ vary according to
\begin{align}
  \dot{\mathscr{P}}_{j m} \vert_{\rm B} & = i \frac{m}{2} \frac{g_{\rm e} \mu_{\rm B}}{\hbar} B \mathscr{P}_{j m} \mbox{,} \label{eq:precession}
\end{align}
where $B$ is the local value of the magnetic field.

\subsection{Spin-exchange collisions}
\label{subsec:collisions}

The spin-polarization of primordial atomic gas can be modified by either spin-exchange collisions or by magnetic interactions \cite{Zygelman05, Zygelman10}. The former dominates under primordial conditions \cite{Zygelman05} so we focus on this process here. The rate coefficients and the resulting evolution of level populations have been calculated by Ref.~\cite{Zygelman05} for a range of temperatures. In order to translate their results into evolution equations for the density matrix, we choose a basis where the density matrix is diagonal. This works because different irreducible moments $\mathscr{P}_{jm}$ of the atomic density matrix do not mix due to collisions in linear theory. Schematically,
\begin{equation}
  \dot{\mathscr{P}}_{j m} \rvert_{\rm c} \sim C_j \mathscr{P}_{j m} \mbox{.} \label{eq:collisionsymm}
\end{equation}
The collision coefficients $C_j$ depend only on the rank of the polarization moment $j$, and not on its projection $m$. Therefore, we can compute the $C_j$ by considering only cases where the $\mathscr{P}_{jm}$ with $m = 0$ are nonzero, i.e. where $\rho_{m_1m_2}$ is diagonal [this follows from Eq.~(\ref{eq:polmom2}) and the $3-j$ symbol selection rule $m=m_2-m_1$]. The equations for the scalar components are slightly more complicated because there are two rank-zero objects that come into play, the occupancies of the singlet and the triplet.\footnote{Alternatively, the collisional evolution of a general density matrix has been studied earlier in Ref.~\cite{Balling64}.}

In such a coordinate system, the rate equations in Ref.~\cite{Zygelman05} take the form:
\begin{subequations}
  \begin{align}
\!\!
\!\!\!\!
    \dot{\rho}_{a a} \rvert_{\rm c} =& -3 k_{\rm x}^+ n_{\rm H} \rho_{a a}^2 + 2(k^- + k^+) n_{\rm H} \rho_{b b} \rho_{d d} \notag \\
    & + 2 k^- n_{\rm H} ( \rho_{b b} + \rho_{d d} ) \rho_{c c}  + k^+ n_{\rm H} \rho_{c c}^2 \notag \\
    & - 2 k_x^- n_{\rm H} \rho_{a a} (\rho_{b b} + \rho_{c c} + \rho_{d d}) \mbox{,} \\
\!\!
\!\!\!\!
    \dot{\rho}_{b b} \rvert_{\rm c} =& \dot{\rho}_{d d} \rvert_{\rm c} = k_{\rm x}^+ n_{\rm H} \rho_{a a}^2  + 2 k_{\rm x}^- n_{\rm H} \rho_{a a}\rho_{c c} + k^0 n_{\rm H} \rho_{c c}^2 \notag \\
    & - (k^0 + k^+ + 2k^-) n_{\rm H} \rho_{b b} \rho_{d d}  \mbox{,~and} \\
\!\!
\!\!\!\!
    \dot{\rho}_{c c} \rvert_{\rm c} =& \, k_{\rm x}^+ n_{\rm H} \rho_{a a}^2 + 2 k_{\rm x}^- n_{\rm H} \rho_{a a} (\rho_{b b} - \rho_{c c} +\rho_{d d}) \notag \\
    & - 2 k^- n_{\rm H} ( \rho_{b b} + \rho_{d d} ) \rho_{c c} - (k^+ + 2k^0) n_{\rm H} \rho_{c c}^2 \notag \\
    & + 2 (k^- +k^0) n_{\rm H} \rho_{b b} \rho_{d d} \mbox{,}
\end{align}
\label{eq:rhodotcollisionraw}
\end{subequations}
where $k^{\pm}$ and $k_x^{\pm}$ represent the thermally averaged (de)excitation rates computed in Ref.~\cite{Zygelman05}.
\supdet{
obtained by averaging the spin-exchange cross sections over a Maxwellian velocity distribution. The latter are given by
\begin{equation}
  k_{\rm x}^{\pm} = \exp{\left(-\omega^{\pm}\right)}k^{\pm} \ \mbox{.} \label{eq:kxpm}
\end{equation}
The factors of $\omega^{\pm}$ are functions of the kinetic temperature, $T_{\rm k}$:
\begin{subequations}
\begin{align}
  \omega^+ & = \frac{2\Delta E_{\rm hf}}{k_{\rm B} T_{\rm k}} = \frac{2 T_*}{T_{\rm k}} \mbox{,} \\
  \omega^- & = \frac{\Delta E_{\rm hf}}{k_{\rm B} T_{\rm k}} = \frac{T_*}{T_{\rm k}} \mbox{.}
\end{align}
\label{eq:omegapm}
\end{subequations}}
We write the evolution equations in \eqref{eq:rhodotcollisionraw} in terms of the irreducible moments $\mathscr{P}_{jm}$ following the argument leading to Eq.~\eqref{eq:collisionsymm} and the explicit forms of Eq.~\eqref{eq:polmom2}. The resulting equations simplify greatly when we assume that the degree of spin-polarization is small, i.e. $\rho_{I I} = \rho_{I I}^{\rm th} + \epsilon_I$ with the thermal populations of the sublevels $\rho_{b b}^{\rm th}=\rho_{c c}^{\rm th}=\rho_{d d}^{\rm th} = \mathscr{P}_{0 0}^{\rm th}/3, \rho_{aa}^{\rm th}=1-\mathscr{P}_{00}^{\rm th}$.
\begin{align}
  \dot{\rho}_{aa} \rvert_{\rm c} & = - n_{\rm H} \kappa(0 \mhyphen 1) \rho_{aa} + n_{\rm H} \kappa(1 \mhyphen 0) \mathscr{P}_{0 0} \mbox{,} \\
  \dot{\mathscr{P}}_{0 0} \rvert_{\rm c} & = n_{\rm H} \kappa(0 \mhyphen 1) \rho_{aa} - n_{\rm H} \kappa(1 \mhyphen 0) \mathscr{P}_{0 0} \notag \\
  & = n_{\rm H} \kappa(0 \mhyphen 1) - n_{\rm H} \left[ \kappa(0 \mhyphen 1) + \kappa(1 \mhyphen 0) \right] \mathscr{P}_{0 0} \mbox{,} \label{eq:p00dotcoll} \\
  \dot{\mathscr{P}}_{1 m} \rvert_{\rm c} & = - n_{\rm H} \kappa^{(1)}(1 \mhyphen 0) \mathscr{P}_{1 m} \mbox{,~and} \\
  \dot{\mathscr{P}}_{2 m} \rvert_{\rm c} & = - n_{\rm H} \kappa^{(2)}(1 \mhyphen 0) \mathscr{P}_{2 m} \mbox{,} \label{eq:p2mdotcoll}
\end{align}
where the coefficients are extensions of the notation of Ref.~\cite{Zygelman05} to include both transition and depolarization rates. Under the simplifying assumptions stated above, these rates are
\begin{subequations}
  \label{eq:collisionrates}
  \begin{align}
    \kappa(0 \mhyphen 1) & = 6 k_{\rm x}^+ (1 - \mathscr{P}_{00}^{\rm th}) + 2 k_{\rm x}^-\mathscr{P}_{00}^{\rm th} \mbox{,} \\
    \kappa(1 \mhyphen 0) & = -2 k_{\rm x}^- (1 - \mathscr{P}_{00}^{\rm th}) + \left( 4 k^- + 2 k^+ \right) \frac{\mathscr{P}_{0 0}^{\rm th}}{3} \mbox{,}\!\! \\
    \kappa^{(1)}(1 \mhyphen 0) & = 0 \mbox{,~and} \\
    \kappa^{(2)}(1 \mhyphen 0) & = 4 k_{\rm x}^- (1 - \mathscr{P}_{00}^{\rm th}) + \frac{2}{3} \left( 3 k^0 + 2 k^- + k^+ \right) \mathscr{P}_{0 0}^{\rm th} \mbox{.}
  \end{align}
\end{subequations}
The depolarization rate $\kappa^{(1)}(1 \mhyphen 0)$ vanishes because the total spin angular momentum of the ensemble, corresponding to the orientation vector $\mathscr{P}_{1 m}$, is conserved in collisions. 

If the spin temperature is much larger than $T_* = 68$ mK, the states are nearly equally occupied and $\mathscr{P}_{00}^{\rm th} \approx \frac34$ [see Eq.~\eqref{eq:spintemperature}]. We use this in the rates of Eq.~\eqref{eq:collisionrates}, along with the expressions for $k_x^\pm$ in terms of $k^\pm$ \cite{Zygelman05}, and write down the collisional contributions to the evolution of the relevant pieces of the atom density matrix:
\begin{subequations}
  \label{eq:atomdmevolcoll}
  \begin{align}
    \dot{\mathscr{P}}_{0 0} \rvert_{\rm c} & = - 4 n_{\rm H} \kappa(1 \mhyphen 0) \left( \mathscr{P}_{0 0} - \frac{3}{4} + \frac{3 T_*}{16 T_{\rm k}} \right) \mbox{~and} \label{eq:p00evolcoll} \\
    \dot{\mathscr{P}}_{2 m} \rvert_{\rm c} & = - n_{\rm H} \kappa^{(2)}(1 \mhyphen 0) \mathscr{P}_{2 m} \mbox{,} \label{eq:p2mevolcoll}
  \end{align}
\end{subequations}
with
\begin{equation}
\kappa^{(2)}(1 \mhyphen 0) = 4\kappa(1 \mhyphen 0) = 2(k^+ + k^-).
\end{equation}
These equations assume that the kinetic temperature $T_{\rm k} \gg T_*$, which is valid over the entire range of redshifts.

\subsection{Optical pumping by Lyman-\texorpdfstring{$\alpha$}{alpha} photons}
\label{subsec:opticalpumping}

Optical pumping by Lyman-$\alpha$ (Ly$\alpha$) photons, or the Wouthuysen-Field effect, is another process that significantly affects the level populations within the hydrogen ground state (see e.g. Ref.~\cite{Happer72}). An atom in the ground ($1{\rm s}$) state absorbs a Ly$\alpha$ photon and gets excited to the $2{\rm p}$ state. Subsequently, the atom reemits a photon and returns to the ground state. However it does not necessarily deexcite to the same ground-state level it originated from. Thus, interactions with Ly$\alpha$ photons can change the density matrix of hydrogen atoms within the ground state basis.

\begin{figure}[t]
\begin{center}
  \includegraphics[width=8cm]{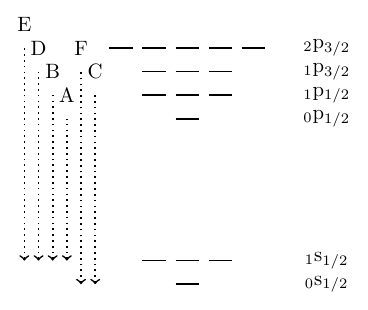}
  \caption{\label{fig:lyman_levels} The hyperfine structure of the ground and first excited electronic levels of the hydrogen atom. The levels are labeled by term symbols $_F l_J$, where $l$ is the spectroscopic notation for the orbital angular momentum, and $J$ and $F$ are the net electronic and total angular momentum respectively. Also shown are all the allowed single photon transitions between the $1{\rm s}$ and $2{\rm p}$ levels, along with their labels; these involve photons in the Ly$\alpha$ frequency range. Only the downward transitions are shown. The gaps between the levels are not drawn to scale.}
\end{center}
\end{figure}

The excited state consists of four levels: $_0{\rm p}_{1/2}$, $_1{\rm p}_{1/2}$, $_1{\rm p}_{3/2}$ and $_2{\rm p}_{3/2}$, where we use the notation $_F l_J$ for the state in terms of its quantum numbers (see Fig.~\ref{fig:lyman_levels}). We use Greek indices represent the excited levels i.e., those within $2{\rm p}$, when used as state labels.

The evolution of the ground state density matrix due to these transitions is governed by second-order perturbation theory, and includes both contributions due to depopulation of $1s$ (Ly$\alpha$ absorption) and repopulation (the subsequent emission). The interaction Hamiltonian between the atom and radiation is
\begin{equation}
H_{{\rm Ly}\alpha}^{{\rm int}} (t) = \!\!\sum_{I, \mu, \alpha, \bm{k}_\gamma} \!\! Q_{\mu I}(\bm{k}_\gamma, \alpha) \vert \mu \rangle \langle I \vert a_\alpha(\bm{k}_\gamma) + \mbox{H.c.,}
\end{equation}
where the matrix element $Q_{\mu I}(\bm{k}_\gamma, \alpha)$ is given by
\begin{equation}
Q_{\mu I}(\bm{k}_\gamma, \alpha) = -i \sqrt{2\pi \hbar \omega} \langle \mu \vert \bm{d}\cdot \bm{e}_\alpha (\hat{\bm{k}}_\gamma)\vert I\rangle \ \mbox{.} \label{eq:lyadipole}
\end{equation}
Here $\bm{d}$ is the electric dipole moment of the atom, which is proportional to the position vector $\bm r$ of the electron. The frequency corresponding to the energy difference between the upper ($\mu$) and lower ($I$) state is $\omega_{\mu I} = (E_\mu - E_I)/\hbar$. The repopulation equation is a straightforward extension of Fermi's golden rule to second order; we have followed the derivation of Eq.~(III,11) of Ref.~\cite{Barrat61}, but with (i) our normalization conventions; (ii) without performing the integration over the outgoing photon wave-vector; and (iii) without averaging over initial photon frequencies.\footnote{We have introduced the latter two modifications because we have multiple excited levels that can interfere with each other, since the hyperfine splitting and natural broadening of the 2p levels are of the same order of magnitude. Note that Eq.~(\ref{eq:repopulation}) could be viewed diagrammatically as the amplitude for the transition $|L\rangle\rightarrow |\nu\rangle\rightarrow|J\rangle$ (including a ``propagator'' for state $|\nu\rangle$), multiplied by the complex conjugate of that for $|K\rangle\rightarrow |\mu\rangle\rightarrow|I\rangle$ since we are following a density matrix rather than an amplitude.} This yields:
\begin{align}
~~~ & \!\!\!\!
\dot{\rho}_{I J} \vert_{\rm repop} \nonumber \\
& = \hbar^{-4}\!\! \!\!\!\! \sum_{\bm k_\gamma, \bm k^\prime_\gamma, \alpha, \beta,\mu, K, \nu, L} \!\!\!\! \!\! f(\bm{k}_\gamma) \ Q^*_{\mu I}(\bm k^\prime_\gamma, \beta) Q_{\mu K}(\bm k_\gamma, \alpha)  \nonumber \\
& ~~~\times  \frac{ Q_{\nu J}(\bm k^\prime_\gamma, \beta) Q^*_{\nu L}(\bm k_\gamma, \alpha)\,\pi \delta(\omega^\prime -\omega+\omega_{JL})  }{\left[ i(\omega-\omega_{\mu K}) - \Gamma_{2{\rm p}}/2 \right]\left[-i(\omega-\omega_{\nu L}) - \Gamma_{2{\rm p}}/2 \right]} \rho_{KL} \nonumber \\
& ~~~+ {\rm c.c.s} \mbox{.} \label{eq:repopulation}
\end{align}
The symbol $\bm k_\gamma$ is the absorbed photon's wave-vector, $\omega$ its frequency, $\alpha$ its polarization index ($\bm k'_\gamma$, $\omega'$, and $\beta$ are used for the reemitted photon). The phase-space density of photons is denoted by $f(\bm{k}_\gamma)$, and $\Gamma_{2{\rm p}}$ is the Einstein $A$-coefficient of the Ly$\alpha$ transition.

We can use Eq.~\eqref{eq:repopulation} to infer a cross-section for the $KL^{\rm th}$ component of the density matrix $\rho$ to transition into the $IJ^{\rm th}$ component. We simplify Eq.~\eqref{eq:repopulation} by using Eq.~\eqref{eq:lyadipole} for the dipole matrix elements, and approximating the incident Ly$\alpha$ radiation field to be isotropic for performing integrals over the directions $\hat{\bm k}_\gamma$ and $\hat{\bm k}^\prime_\gamma$. This is an excellent approximation due to the large value of the cross-section, and low mean free path for incident Ly$\alpha$ photons. Thus we conclude that Eq.~\eqref{eq:repopulation} connects only irreducible components of the same rank within the initial and final density matrix.

Let the initial and final states, ($I$ and $J$), belong to multiplets with total angular momentum quantum numbers $F_I$ and $F_J$ respectively.  We implement the above program to infer the cross-section for a general irreducible component of rank-$j$ within the initial state submatrix to go to the corresponding component within the final state submatrix. We use the suggestive notation $\sigma_{F_I \rightarrow F_J, (j)}$ to represent this cross-section; for example, the contribution to the density matrix from $F=1\rightarrow 1$ scattering is
\begin{equation}
  \dot{\mathscr{P}}_{jm} \rvert_{{\rm repop},\,F=1\rightarrow 1}  = 4\pi \int {\rm d}\nu\, J_{{\rm Ly}\alpha} (\nu)
  \sigma_{1 \rightarrow 1, (j)} (\nu)\mathscr{P}_{j m}
  \mbox.
\end{equation}
The expression for $\sigma_{F_I \rightarrow F_J, (j)}$ can be read off from Eq.~\eqref{eq:repopulation}.
We approximate all multiplicative factors of frequencies by the value of the Ly$\alpha$ line-center, and get (using e.g. the methodology of Ref.~\cite{Hirata06})
\begin{align}
~~ & \!\!\!\!
  \sigma_{F_I \rightarrow F_J, (j)} (\omega) \nonumber \\
  & = \frac{ 8 \pi }{9} \frac{ \omega_{{\rm Ly}\alpha}^4 }{ c^4 } \frac{ e^4 }{ \hbar^2} \sqrt{2j+1} \sqrt{ \frac{2F_J + 1}{2F_I + 1} } \sum_{m_{J_1}, m_{J_2}} \sum_{m_{I_1}, m_{I_2}} \nonumber \\
  & ~~~ \sum_{j^\prime, m^\prime} \sum_{p, q} \sum_{r, s} \sum_{\mu, \nu} (-1)^{F_J - m_{J_2}} \tj{F_J}{j}{F_J}{-m_{J_2}}{m}{m_{J_1}} \nonumber \\
  & ~~~ \times (-1)^{F_I - m_{I_2}} \tj{ F_I }{ j^\prime }{ F_I }{-m_{I_2}}{ m^\prime }{m_{I_1}} g_{p r} g_{q s} \nonumber \\
  & ~~~ \times \frac{ \langle F_J m_{J_1} \vert r^p \vert \mu \rangle \langle \mu \vert r^q \vert F_I m_{I_1} \rangle  }{ \Delta \omega_{\mu I} + i \Gamma_{\mu}/2  }
\nonumber \\ &~~~ \times
\frac{\langle F_I m_{I_2} \vert r^s \vert \nu \rangle \langle \nu \vert r^r \vert F_J m_{J_2} \rangle}{\Delta \omega_{\nu I} - i \Gamma_{\nu}/2}
   \mbox{,} \label{eq:cross-section-Lya-unsimpl}
\end{align}
where $\Delta \omega_{\mu I} = \omega - \omega_{\mu I}$ is the frequency offset from the line-center of the $\mu \rightarrow I$ transition, and $g_{pr} = (-1)^p\delta_{p,-r}$ is the metric tensor. The two 3-j symbols project the irreducible components of rank $j$ and $j^\prime$ (which equals $j$) in the initial and final density submatrices in Eq.~\eqref{eq:repopulation}. We further simplify this result using the Wigner-Eckart theorem, and collapse the sum of six 3-j symbols:
\begin{align}
  \sigma_{F_I \rightarrow F_J, (j)} (\omega)
  & = \frac{ 8 \pi }{9} \frac{ \omega_{{\rm Ly}\alpha}^4 }{ c^4 } \frac{ e^4 }{ \hbar^2} \sqrt{ \frac{ 2 F_J + 1 }{ 2 F_I + 1 }  } \sum_{\mu, \nu} (-1)^{F_I - F_J} \nonumber \\
  & ~~~ \times \frac{ \langle \mu \| r \| J \rangle^\ast \langle \mu \| r \| I \rangle \langle \nu \| r \| I \rangle^\ast \langle \nu \| r \| J \rangle }{ (\Delta \omega_{\mu I} + i \Gamma_\mu/2 )(\Delta \omega_{\nu I} - i \Gamma_\nu/2 ) } \nonumber \\
  & ~~~ \times \sj{F_{\mu}}{F_{\nu}}{j}{F_I}{F_I}{1} \sj{F_{\mu}}{F_{\nu}}{j}{F_J}{F_J}{1} \mbox{.} \label{eq:cross-section-Lya}
\end{align}
When we perform the summation over the upper levels ($\mu$ and $\nu$), the terms with $\mu = \nu$ and $\mu \ne \nu$ give Lorentzian line and interference profiles respectively. In this calculation, we assumed that the only factor involved in broadening the lines shown in Fig.~\ref{fig:lyman_levels} is their finite lifetime; in reality, the lines are broadened due to a combination of this and the Doppler effect, owing to which we need to convolve these profiles with the appropriate velocity distributions.

In the case where the triplet sublevels are equally occupied, the only relevant components of the density submatrices are those of rank zero. For $j=0$, Eq.~\eqref{eq:cross-section-Lya} gives net transition cross-sections from $F=1 \rightarrow F=0$ and $F = 0 \rightarrow F=1$, which have been previously worked out. We use the notation and list of line strengths in Appendix B of Ref.~\cite{Hirata06}. In particular, Fig.~\ref{fig:lyman_levels} shows their choice of labels for the various lines making up the fine-structure of the Ly$\alpha$ line, which we will use in subsequent expressions. 

Using the line-strengths in Ref.~\cite{Hirata06} for the irreducible matrix elements in Eq.~\eqref{eq:cross-section-Lya}, the isotropic cross-sections $\sigma_{F_I\rightarrow F_J,(0)}$ are given by Eqs.~(B17,B18) of Ref.~\cite{Hirata06}.The one new cross section we need is the rank-2 cross section,
\begin{align}
  \sigma_{1 \rightarrow 1, (2)} 
  & = \frac32 \lambda_{{\rm Ly}\alpha}^2 \gamma_{2{\rm p}} \Bigl( \frac{1}{27} \phi_{\rm BB} + \frac{1}{108} \phi_{\rm DD} + \frac{7}{36} \phi_{\rm EE}  \nonumber \\
  & ~~~ + \frac{2}{9} \phi_{\rm AE}+ \frac{1}{27} \phi_{\rm BD} + \frac{1}{3} \phi_{\rm BE} + \frac{1}{6} \phi_{\rm DE} \Bigr) \mbox{,} \label{eq:alignmentrepop}
\end{align}
where $\gamma_{2{\rm p}} = \Gamma_{2{\rm p}}/4\pi = 50 \ {\rm MHz}$ and $\phi_{\rm AB}$ etc.\ are the Lorentzian profiles defined by Eq.~(B16) of Ref.~\cite{Hirata06}.

We also need the depopulation rates (or equivalent cross-section). These rates are independent of the rank of the irreducible component (or the magnetic quantum numbers) because of the isotropy of the Lyman-$\alpha$ radiation -- all of the 1s $F=1$ states are depopulated at the same rate. Since the net population of the $2{\rm p}$ levels is always negligible, the rate of depopulation from a level is given by the sum of the rates of all repopulations which start from that level:
\begin{align}
  \sigma_{F_I} \vert_{\rm depop} & = \sum_{F_J} \sigma_{F_I \rightarrow F_J, (0)} \mbox{.}
\end{align}
We obtain the following evolution equations for the irreducible components of interest by subtracting the contribution of depopulation from that of repopulation:
\begin{align}
  \dot{\mathscr{P}}_{0 0} \rvert_{{\rm Ly}\alpha} & = - 4 \pi \int {\rm d}\nu J_{{\rm Ly}\alpha} (\nu) \Bigl[ \sigma_{1 \rightarrow 0, (0)} (\nu) \mathscr{P}_{0 0} \nonumber \\
  & \hspace{40pt} - \sigma_{0 \rightarrow 1, (0) } (\nu) \rho_{aa} \Bigr] \mbox{,} \label{eq:p00dotlya_raw} \\
  \dot{\mathscr{P}}_{2 m} \rvert_{{\rm Ly}\alpha} & = - 4\pi \int {\rm d}\nu J_{{\rm Ly}\alpha} (\nu) \Bigl[ \sigma_{1 \rightarrow 1, (0)} (\nu) + \sigma_{1 \rightarrow 0, (0)} (\nu) \nonumber \\
  & \hspace{40pt} - \sigma_{1 \rightarrow 1, (2)} (\nu) \Bigr] \mathscr{P}_{2 m} \label{eq:p2mdotlya_raw} \mbox{.}
\end{align}
To simplify these equations, we use the relation $\rho_{aa} = 1 - \mathscr{P}_{0 0}$, and substitute the repopulation cross-sections.

The effect of optical pumping by Ly$\alpha$ photons on the rank zero component (net triplet occupancy) is complicated by a source term. In the approximation of a very high cross-section (or $T>\infty$), the states are driven to equal occupancy i.e. $\mathscr{P}_{0 0} \rightarrow 3/4$. In order to correct the populations for a finite temperature, we need to consider the frequency dependence of the flux $J_{{\rm Ly}\alpha}$. This motivates the definition of the flux correction factor $\tilde{S}_{\alpha}$ \footnote{The tilde is to avoid conflict with the usual definition of $S_{\alpha}$ in the literature, which approximates the color temperature, $T_{\rm c, eff}$ with the kinetic temperature, $T_{\rm k}$. It is consistent with the notation of Ref.~\cite{Hirata06}.} and the effective color temperature $T_{\rm c, eff}$, which are given by
\begin{equation}
  T_{\rm c, eff}  = - \frac{ h }{ k_{\rm B} } \frac{{\rm d} }{{\rm d} \nu} \ln{J}_{{\rm Ly}\alpha} (\nu)  
\end{equation}
and
\begin{equation}   
  \tilde{S}_{\alpha}  = \frac{9}{8 \lambda_{{\rm Ly}\alpha}^2 \gamma_{2{\rm p}} } \int {\rm d}\nu \frac{ J_{{\rm Ly}\alpha} (\nu) }{J_\alpha}  [ \sigma_{1 \rightarrow 0, (0)} (\nu) + \sigma_{0 \rightarrow 1, (0)} (\nu) ] \mbox{,}
\end{equation}
where $J_{\alpha}$ is the flux on the blue side of the Lyman-$\alpha$ line, before it is processed by any radiative transfer. Substitution of these definitions in Eq.~\eqref{eq:p00dotlya_raw} gives us the evolution equation for the occupancy:
\begin{align}
  \dot{\mathscr{P}}_{0 0} \rvert_{{\rm Ly}\alpha} & = - \frac{32}{9} \pi \lambda_{{\rm Ly}\alpha}^2 \gamma_{2{\rm p}} \tilde{S}_{\alpha} J_{\alpha} \Bigl[ \mathscr{P}_{0 0} - \frac34 + \frac{3 T_*}{16 T_{\rm c, eff} } \Bigr] \mbox{.} \label{eq:p00dotlya}
\end{align}
The evolution of the rank two irreducible component of the triplet state density submatrix is easier to evaluate, since it has no source term. The detailed frequency dependence of the flux $J_{{\rm Ly}\alpha}$ is not crucial. Substituting the expressions for the cross-sections, we obtain the following depolarization rate:
\begin{align}
  \dot{\mathscr{P}}_{2 m} \rvert_{{\rm Ly}\alpha} & = - 0.601 \times 6 \pi \lambda_{{\rm Ly}\alpha}^2 \gamma_{2{\rm p}} \tilde{S}_{\alpha, (2)} J_{\alpha} \mathscr{P}_{2 m} \mbox{,} \label{eq:p2mdotlya}
\end{align}
where the flux correction factor $\tilde{S}_{\alpha, (2)}$ for the rank-two tensor is defined such that
\begin{multline}
\!\!\!\!\!\!
  0.601 \tilde{S}_{\alpha, (2)} J_{\alpha} = \int {\rm d} \nu J_{{\rm Ly} \alpha}(\nu) \Bigl( \frac19 \phi_{\rm AA} + \frac{5}{27} \phi_{\rm BB} + \frac{11}{108} \phi_{\rm DD} \\
   + \frac{13}{36} \phi_{\rm EE}  - \frac{2}{9} \phi_{\rm AE} - \frac{1}{27} \phi_{\rm BD} - \frac13 \phi_{\rm BE} - \frac16 \phi_{\rm DE} \Bigr) \mbox{,}
\end{multline}
and the numerical prefactor is the integral over frequency of the term enclosed in braces on the RHS of the above equation.

\section{Radiative Transfer}
\label{sec:radtrans}

Sections \ref{sec:dmevolnradiative} and \ref{sec:otherproc} dealt with the evolution of the atom's density matrix due to various processes. In this section, we study the evolution of the components of the $21$-cm radiation's phase-space density matrix $f_{X,jm}(\omega)$. In particular, the intensity monopole $f_{\text{I},0 0}$ and quadupole $f_{\text{I}, 2 m}$ are the relevant multipoles to study for the effect on the brightness temperature.

The baryon rest frame simplifies the details of the matter-radiation interaction, hence we use it throughout this calculation. We restrict ourselves to quantities which are at most of the first order in smallness in terms of the matter overdensity $\delta$.

The only quantity related to the radiation field with a zeroth-order piece is the intensity monopole $f_{\text{I}, 0 0}$. From the discussion in Sec.~\ref{sec:introduction}, we expect the matter velocity $\bm v$ and the intensity and polarization quadrupoles, $f_{\text{I}, 2 m}$ and $f_{E, 2 m}$, to be quantities of the first order in smallness.

The Boltzmann equation for a generic component of the phase space density $f_{X}$ is
\begin{equation}
  \frac{{\rm D} f_X}{{\rm D} t} = \dot{f}_X \rvert_{\rm{s}} \mbox{.} \label{eq:boltzmann}
\end{equation}
The left-hand side is the material derivative with respect to the flow of points in phase space, which represents the effect of free-streaming. The right-hand side is the source term for the phase-space density, due to interaction with atoms. 

\subsection{Free-streaming term}
\label{subsec:freestream}

The material derivative of the phase-space density expands to
\begin{equation}
  \frac{{\rm D} f}{{\rm D} t} = \dot{f} + \frac{{\rm d} \bm x}{{\rm d} t} \cdot \bm \nabla f + \frac{{\rm d} \omega}{{\rm d} t} \frac{\partial f}{\partial \omega} + \frac{{\rm d} \hat{\bm n}}{{\rm d} t} \cdot \nabla_{\hat{\bm n}} f \mbox{,} \label{eq:freestreaming}
\end{equation}
where, as earlier, $\hat{\bm n}$ is the radiation's direction of propagation. The second, third, and fourth terms represent advection, time-dependent redshift, and lensing, respectively. Since we are interested only in terms up to the first order in the density fluctuations, we neglect lensing (since it is a second-order effect), and replace the coefficient of $\bm \nabla f$ in the advection term with its zeroth-order value, which is
\begin{equation}
  \frac{{\rm d} \bm x}{{\rm d} t} = c \, \hat{\bm n} \mbox{.} \label{eq:velexpand}
\end{equation}
In order to expand the redshift term, we use the relation between the angular frequency of a photon in the baryon rest frame ($\omega$) and in the Newtonian frame ($\omega_{\rm N}$):
\begin{equation}
  \omega = \omega_{\rm N} \left( 1 - \frac{\bm v \cdot \hat{\bm n}}{c} \right) \mbox{,} \label{eq:dopplershift} 
\end{equation}
where $\bm v$ and $\hat{\bm n}$ are the bulk matter velocity and the direction of the photon's travel respectively. The coefficient of the time-dependent redshift term is
\begin{align}
  \frac{1}{\omega} \frac{{\rm d} \omega}{{\rm d} t} & = \frac{1}{\omega_{\rm N}} \frac{{\rm d} \omega_{\rm N}}{{\rm d} t} - \frac{1}{c} \frac{{\rm d}}{{\rm d} t}(\bm v \cdot \hat{\bm n}) + \hdots \notag \\
  & = \frac{1}{\omega_{\rm N}} \frac{{\rm d} \omega_{\rm N}}{{\rm d} t} - \frac{1}{c} \dot{v}_i n_i - \frac{\partial v_i}{\partial x_j} n_i n_j + \hdots
\end{align}
The first term, which is the rate of redshifting in the Newtonian frame, has contributions both from large-scale Hubble flow and gravitational redshifting in the presence of local potential wells \cite{1995ApJ...455....7M}. The latter contribution is the Sachs-Wolfe effect. The second term is the time-dependent redshift due to local acceleration, and is of the same size as the Sachs-Wolfe term. The final term, which is the origin of the effect of interest, is the contribution of the local matter velocity gradient $\bm \nabla \bm v$. 

The effect of local velocity gradients is much larger than that of acceleration, which scales as the depth of the potential wells, as long as the modes under consideration are subhorizon sized. We estimate their relative sizes as
\begin{multline}
  \frac{(1/c) \dot{v}_i n_i}{(\partial v_i/\partial x_j) n_i n_j} \approx \frac{a H}{k c} \approx \, 4 \times 10^{-4} \\
  \times \left(\frac{1+z}{10}\right)^{1/2} \left(\frac{k}{1 \text{ Mpc}^{-1}}\right)^{-1} \left(\frac{\Omega_m h^2}{0.143}\right)^{1/2}\mbox{.} \label{eq:dipvsquad}
\end{multline}
The second term in Eq.~\eqref{eq:freestreaming} is the advection term. On free streaming, it causes mixing of multipoles on a characteristic timescale $\sim (a/k c)$ \cite{HuWhite97}. The size of this contribution relative to the time-dependent redshift term is set by the comparison with the timescale for the photons to redshift through the line. We can safely neglect the advection term as long as we restrict ourselves to modes of wavelengths much larger than the Jeans length, $r_{\rm J}$, at this epoch. This is a good approximation for the modes under consideration:
\begin{multline}
  \frac{c (\partial f/\partial x_i) n_i}{H \omega (\partial f/\partial \omega)} \sim \, \frac{k}{a} \frac{v_{\rm s}}{H} \sim \frac{k r_{\rm J}}{a} \approx \, 5.8 \times 10^{-3} \\
  \times \left(\frac{T_{\rm k}}{T_\gamma}\right)^{1/2} \left(\frac{k}{1 \text{ Mpc}^{-1}}\right) \left(\frac{\Omega_{\rm m} h^2}{0.143}\right)^{-1/2} \mbox{.}
\end{multline}
Hence the most important contribution to the time-dependent redshift term is the velocity gradient term. We assume that the fluctuation is a plane wave with comoving wave-vector $\bm k$, and use the continuity equation to express the velocity gradient in terms of the overdensity as follows:
\begin{equation}
  \frac{1}{\omega} \frac{{\rm d} \omega}{{\rm d} t}  \approx - H - \frac{\partial v_i}{\partial x_j} n_i n_j
   = - H \Bigl[1 - \delta (\hat{\bm k} \cdot \hat{\bm n})^2 \Bigr] \mbox{,} \label{eq:dopplerexpand} 
\end{equation}
where $H$ is the Hubble rate at the redshift under consideration and $\delta$ is the local overdensity.  In writing this equation, we used the standard scaling of the growth factor for a matter dominated universe, i.e. $d(\log \delta)/d(\log a) = 1$. 

Thus the free-streaming term of Eq.~\eqref{eq:freestreaming} is
\begin{equation}
  \frac{{\rm D} f}{{\rm D} t} = \dot{f} - H \Bigl[ 1 - \delta (\hat{\bm k} \cdot \hat{\bm n})^2 \Bigr] \omega \frac{\partial f}{\partial \omega} \label{eq:freestreamcartesian} \mbox{.}
\end{equation}
In a coordinate system with an arbitrary orientation,
\begin{equation}
  (\hat{\bm k} \cdot \hat{\bm n})^2 = \frac{8\pi}{15} \sum_m Y_{2 m}(\hat{\bm k}) \left[ Y_{2 m}(\hat{\bm n}) \right]^* + \frac{1}{3} \mbox{.} \label{eq:mu2}
\end{equation}
Using this identity, we write down the free-streaming terms for the relevant moments in a general coordinate system. 

In order to expand Eq.~\eqref{eq:freestreamcartesian} into moments, we note that the only relevant moments i.e. those which are nonzero up to first order in the matter density fluctuation $\delta$, are the intensity monopole $f_{\text{I},0 0}$ (which has a zeroth-order piece too) and quadrupole $f_{\text{I}, 2 m}$, and the polarization quadrupole $f_{{\rm E}, 2 m}$ (vide Sec.~\ref{sec:estimate} and Table \ref{tab:sizes}). Thus, up to first order in $\delta$, the equations describing the free-streaming of the relevant moments are
\begin{subequations}
  \label{eq:photonfreestream}
  \begin{align}
    \frac{{\rm D} f_{\text{I}, 0 0}}{{\rm D} t} & = \dot{f}_{\text{I}, 0 0} - H \left[ 1 - \frac{ \delta }{3} \right] \omega \frac{\partial f_{\text{I}, 0 0}}{\partial \omega} \mbox{,} \\
    \frac{{\rm D} f_{\text{I}, 2 m}}{{\rm D} t} & = \dot{f}_{\text{I}, 2 m} - H \omega \frac{\partial f_{\text{I}, 2 m}}{\partial \omega} 
    \nonumber \\ & ~~~~+ \frac{2}{3} \sqrt{\frac{4\pi}{5}} \delta H \omega \frac{\partial f_{\text{I}, 0 0}}{\partial \omega} Y_{2 m}(\hat{\bm k}) \mbox{,~and} \\
    \frac{{\rm D} f_{{\rm E}, 2 m}}{{\rm D} t} & = \dot{f}_{{\rm E}, 2 m} - H \omega \frac{\partial f_{E, 2 m}}{\partial \omega}  \mbox{.}
  \end{align}
\end{subequations}

\subsection{Source term}
\label{subsec:collisionterm}

The source term describes the evolution of the $21$-cm radiation's phase-space density matrix due to interaction with neutral hydrogen atoms. In this section, we generalize the usual treatment of spontaneous and stimulated emission, and photo-absorption to the case of spin-polarized atoms.

We complete construction of the plane wave source term $\dot f_{\alpha\beta}(\hat{\mathbf n},\omega) \rvert_{\rm{s}}$ in several steps. First, we find the contribution to the plane wave source term from a single atom in terms of spherical operators. Then we sum this contribution over all atoms, with the specified number density $n_{\rm H}x_{1{\rm s}}$. Finally, we turn the required expectation values of spherical operators into photon phase space densities, and reexpress them in terms of the radiation multipoles and atomic polarizations.

We write the second-order moments of the photon field in the plane wave basis in terms of the spherical basis by inversion of Eq.~\eqref{eq:photonwavefunction}:
\begin{equation}
  a_\alpha({\bm k}_\gamma) = \frac{(2\pi c)^{3/2}}{\omega} e^{-i{\bm k}_\gamma \cdot{\bm R}}\sum_{jm\lambda} \left[ {\bm e}^\ast_{(\alpha)}\cdot {\bm Y}_{jm}^{(\lambda)}\right](\hat{\bm k}_\gamma) \, a_{jm}^{(\lambda)}(\omega) \mbox{,}
\end{equation}
where $\omega = k_\gamma/c$ and $\lambda\in\{{\rm E}, {\rm M}\}$. We have inserted a factor of $e^{-i{\bm k_\gamma}\cdot{\bm R}}$ here to place the atom (which is the center around which we expand the spherical waves) at position ${\bm R}$ rather than the origin. It follows that the time evolution of the photon density matrix is
\begin{align}
~~ & \!\!\!
\frac {\rm d}{{\rm d} t} \langle a_\alpha^\dagger({\bm k}_\gamma) a_\beta({\bm k}^\prime_\gamma) \rangle \notag \\
& = \frac{(2\pi c)^{3}}{\omega^2}\!\!
\sum_{jm\lambda j'm'\lambda'} \!\!\left[ {\bm e}_{(\alpha)}\cdot {\bm Y}_{jm}^{(\lambda)\ast}\right](\hat{\bm k}_\gamma)
\left[ {\bm e}^\ast_{(\beta)}\cdot {\bm Y}_{j'm'}^{(\lambda')}\right](\hat{\bm k}^\prime_\gamma)
\nonumber \\ 
& ~~~\times e^{i({\bm k}_\gamma - {\bm k}^\prime_\gamma)\cdot{\bm R}}
\frac {\rm d}{{\rm d} t} \langle a_{jm}^{(\lambda)\dagger}(\omega)
a_{j'm'}^{(\lambda')}(\omega') \rangle.
\label{eq:temp-chris-1}
\end{align}
This result is valid if the electromagnetic field interacts with a single atom. However, in the scenario under consideration, it interacts with an ensemble of atoms of number density $n_{\rm H}x_{1{\rm s}}$. We obtain such an ensemble by integrating Eq.~(\ref{eq:temp-chris-1}) over volume $d^3{\bm R}$ and multiplying by $n_{\rm H} x_{1{\rm s}}$. Using the rule that $\int e^{i({\bm k}_\gamma - {\bm k}^\prime_\gamma)\cdot{\bm R}} \,d^3{\bm R} = (2\pi)^3 \delta^{(3)}({\bm k}_\gamma-{\bm k}^\prime_\gamma)$, we obtain a $\delta$-function on the right-hand side and hence the result:
\begin{align}
  \dot f_{\beta\alpha}(\omega,\hat{\bm k}_\gamma) \vert_{\rm s}
  & = \frac{(2\pi c)^{3}}{\omega^2} n_{\rm H} x_{1{\rm s}}
  \sum_{jm\lambda j'm'\lambda'} \left[ {\bm e}_{(\alpha)}\cdot {\bm Y}_{jm}^{(\lambda)\ast}\right](\hat{\bm k}_\gamma)
  \nonumber \\ 
  & ~~~\times
  \left[ {\bm e}^\ast_{(\beta)}\cdot {\bm Y}_{j'm'}^{(\lambda')}\right](\hat{\bm k}_\gamma)
  \frac {\rm d}{{\rm d} t} \langle a_{jm}^{(\lambda)\dagger}(\omega)
  a_{j'm'}^{(\lambda')}(\omega) \rangle \mbox{.}
  \label{eq:temp-chris-2}
\end{align}
Note that in Eq.~(\ref{eq:temp-chris-2}), the derivative on the right-hand side is the contribution of a single atom.

Since the operator $a_{jm}^{(\lambda)\dagger}(\omega) a_{j'm'}^{(\lambda')}(\omega)$ commutes with the radiation's Hamiltonian $H_\gamma$, it evolves only in accordance with the interaction Hamiltonian $H_{{\rm hf},\gamma}$, specifically:
\begin{align}
~~ & \!\!\!
\frac {\rm d}{{\rm d} t} \langle a_{jm}^{(\lambda)\dagger}(\omega)
a_{j'm'}^{(\lambda')}(\omega) \rangle
\nonumber \\
& = \frac i\hbar \left\langle \left[ H_{{\rm hf},\gamma}, a_{jm}^{(\lambda)\dagger}(\omega)
a_{j'm'}^{(\lambda')}(\omega) \right]\right\rangle
\nonumber \\
& = \frac i\hbar \sum_{m_F} V_{m_Fa,m}(\omega) \left\langle |1m_F\rangle \langle 00| a^{(\lambda')}_{j'm'}(\omega) \right\rangle
\delta_{{\rm M} \lambda} \delta_{j1}
\nonumber \\ & ~~~~ + {\rm c.c.s.}
\nonumber \\
& = -\frac{\pi}{\hbar^2}\! \sum_{m_F m_2 m_3}\!\! V_{m_Fa,m}(\omega) V^\ast_{m_2a,m_3}(\omega) \delta(\omega-\omega_{\rm hf})
\delta_{{\rm M} \lambda}\delta_{j1}
\nonumber \\ 
& ~~~\times
\Bigl[
\delta_{m_2m_F}\rho_{aa}f_{m'm_3}^{(\lambda'j')({\rm M}1)}(\omega)
- \rho_{m_2m_F}(\delta_{\lambda'{\rm M}}\delta_{j'1}\delta_{m'm} 
\nonumber \\ 
& ~~~~
+ f_{m'm_3}^{(\lambda'j')({\rm M}1)}(\omega))
\Bigr] + {\rm c.c.s.}
\label{eq:temp-chris-3}
\end{align}
Here again ``c.c.s.'' stands for complex conjugation with a swap (i.e. swap $\lambda jm \leftrightarrow \lambda'j'm'$). In the second equality we used Eq.~\eqref{eq:M1Hamilt} for $H_{{\rm hf},\gamma}$, and in the third we use the results of Appendix \ref{sec:3ptfn} for the atom-radiation three-point function.

We next use Eq.~\eqref{eq:vintermsofa} for the interaction matrix elements, with which Eq.~(\ref{eq:temp-chris-3}) simplifies to
\begin{align}
~~ & \!\!\!
\frac {\rm d}{{\rm d} t} \langle a_{jm}^{(\lambda)\dagger}(\omega)
a_{j'm'}^{(\lambda')}(\omega) \rangle
\nonumber \\
& = -\frac{A}{2} \delta(\omega-\omega_{\rm hf})
\delta_{{\rm M}\lambda}\delta_{j1}
\sum_{m_2}
\Bigl\{
\delta_{m_2m}\rho_{aa}f_{m'm_2}^{(\lambda'j')({\rm M}1)}(\omega)
\nonumber \\ 
& ~~~~
- \rho_{m_2m}[\delta_{\lambda'{\rm M}}\delta_{j'1}\delta_{m'm_2} 
+ f_{m'm_2}^{(\lambda'j')({\rm M}1)}(\omega)]
\Bigr\} + {\rm c.c.s.}
\label{eq:fdotspherical}
\end{align}
A useful definition is the isotropic absorption cross-section $\sigma(\omega)$ for radiation whose wavelength is close to $21$-cm:
\begin{equation}
\sigma(\omega) = 3 \pi^2 \frac{c^2}{\omega^2} A \, \phi(\omega) \mbox{,} \label{eq:crosssection}
\end{equation}
where $\phi(\omega)$ is the absorption profile centered at $\omega_{hf}$. It is broadened from the delta function of Eq.~\eqref{eq:fdotspherical} due to the thermal motions of the hydrogen atoms.

Substituting Eq.~\eqref{eq:fdotspherical} into Eq.~(\ref{eq:temp-chris-2}), using the definition \eqref{eq:crosssection} and the notation $\hat{\bm n}$ for the direction of propagation, we get
\begin{align}
~~~~& \!\!\!\!\!\!\!\!\!\!  \dot f_{\beta\alpha}(\omega,\hat{\mathbf n}) \vert_{\rm s}
\nonumber \\
  = & -\frac{4 \pi}{3} n_{\rm H} x_{1{\rm s}} \sigma(\omega) c
  \sum_{m_2m j'm'\lambda'} \left[ {\bm e}_{(\alpha)}\cdot {\bm Y}_{1m}^{({\rm M})\ast}\right](\hat{\mathbf n})
  \nonumber \\
  & \times
  \left[ {\bm e}^\ast_{(\beta)}\cdot {\bm Y}_{j'm'}^{(\lambda')}\right](\hat{\mathbf n})
  \Bigl\{
    \delta_{m_2m}\rho_{aa}f_{m'm_2}^{(\lambda'j')({\rm M}1)}(\omega)
    \nonumber \\ 
    & ~~~~
    - \rho_{m_2m}[\delta_{\lambda'{\rm M}}\delta_{j'1}\delta_{m'm_2} 
      + f_{m'm_2}^{(\lambda'j')({\rm M}1)}(\omega)]
    \Bigr\}
  \nonumber \\ 
  & ~~~~
  + [\alpha\leftrightarrow\beta]^\ast.
  \label{eq:temp-chris-4}
\end{align}
(Note that because of the symmetry of Eq.~\ref{eq:temp-chris-2} under $\lambda jm\leftrightarrow\lambda'j'm'$ symmetry, the ``c.c.s.'' term simply results in the complex conjugate of the contribution with $\alpha$ and $\beta$ switched, thereby guaranteeing the Hermiticity of the phase-space density matrix.)

It is profitable to break Eq.~(\ref{eq:temp-chris-4}) into the three terms in braces: these correspond to absorption, spontaneous emission, and stimulated emission, respectively. Each one may be converted back into radiation multipole moments using the inverse of Eq.~\eqref{eq:momentexpansion}:
\begin{equation}
  (\dot f_{\beta\alpha})_{jm} (\omega) \vert_{\rm s} = \sqrt{\frac{2j+1}{4\pi}} \int \dot f_{\beta\alpha}(\omega,\hat{\bm n}) \vert_{\rm s} [{}_{\beta-\alpha}Y_{jm}(\hat{\bm n})]\,d^2{\bm n}.
\end{equation}
This conversion entails the angular integral of products of three spherical harmonics, and results in appropriate sets of 3-j symbols \cite{Varshalovich88}.

The absorption term is
\begin{equation}
  (\dot{f}_{\alpha \beta})_{j m} (\omega) \vert_{\rm{ab}} = - n_{\rm H} x_{1{\rm s}} \sigma(\omega) c \, \rho_{a a} \, (f_{\alpha \beta})_{j m} (\omega) \mbox{.} \label{eq:photoabsph}
\end{equation}
The emission terms involve elements of the triplet state density submatrix $\rho_{m n}$, which are most naturally expressed in terms of the irreducible components $\mathscr{P}_{j m}$ using Eq.~\eqref{eq:polmom2}. The spontaneous emission term simplifies to
\begin{align}
  (\dot{f}_{\alpha \beta} )_{j m} (\omega) \vert_{\rm sp.em} & = n_{\rm H} x_{1{\rm s}} \frac{\sigma(\omega) c}{3} \sqrt{3(2j+1)} \notag \\
  & ~~~\times \alpha \beta \tj{1}{1}{j}{\alpha}{-\beta}{\beta-\alpha} \mathscr{P}_{j m} \mbox{,} \label{eq:photonspontem}
\end{align}
and the stimulated emission term simplifies to 
\begin{align}
  ~~ & \!\!\! 
  (\dot{f}_{\alpha \beta})_{j m} \rvert_{\rm st.em} \notag \\
  & = \frac{2j+1}{2} n_{\rm H} x_{1{\rm s}} \frac{\sigma(\omega) c}{3} (-1)^{m} \!\!\! \sum_{j_1 m_1 j_2 m_2 \gamma} \notag \\
  & ~~~\sqrt{3(2j_2 + 1)} 
  \Bigl[ \alpha \gamma 
    \tj{j_1}{j_2}{j}{-m_1}{-m_2}{m} 
    \tj{1}{1}{j_2}{\alpha}{-\gamma}{\gamma - \alpha}
    \notag \\
  & ~~~~~~\times 
  \tj{j_1}{j_2}{j}{\gamma - \beta}{\alpha - \gamma}{\beta - \alpha}
  (f_{\gamma \beta})_{j_1 m_1} \mathscr{P}_{j_2 m_2} \Bigr] \notag \\
  & ~~~~~~+ 
  (-1)^{-m} [\alpha \leftrightarrow \beta, m \rightarrow -m]^\ast \mbox{.}
  \label{eq:photonstimem}
\end{align}
We can further rewrite the source terms of \eqref{eq:photoabsph}, \eqref{eq:photonspontem} and \eqref{eq:photonstimem} in terms of the parity invariants of Eq.~\eqref{eq:photonparitymoments}. 

As noted earlier in Sec.~\ref{subsec:freestream}, the only relevant moments are the intensity monopole $f_{\text{I},0 0}$ and quadrupole $f_{\text{I}, 2 m}$, and the polarization quadrupole $f_{E, 2 m}$. Summing up all the contributions yields the following source terms for these moments:
\begin{subequations}
  \label{eq:photoncollision}
  \begin{align}
    \dot{f}_{\text{I}, 0 0}(\omega) \vert_{\rm s} & = n_{\rm H} x_{1{\rm s}} \frac{\sigma(\omega) c}{3} \left[ - \left(3 - 4 \mathscr{P}_{0 0} \right) f_{\text{I}, 0 0} + \mathscr{P}_{0 0} \right] \mbox{,} \\
    \dot{f}_{\text{I}, 2 m}(\omega) \vert_{\rm s} & = n_{\rm H} x_{1{\rm s}} \frac{\sigma(\omega) c}{3} \bigl[ -\left(3 - 4\mathscr{P}_{0 0} \right) f_{\text{I}, 2 m} \notag \\
      & ~~~+ \frac{1}{\sqrt{2}} \left(1 + f_{\text{I}, 0 0}\right) \mathscr{P}_{2 m} \bigr] \mbox{,~and} \\
    \dot{f}_{{\rm E}, 2 m}(\omega) \vert_{\rm s} & = n_{\rm H} x_{1{\rm s}} \frac{\sigma(\omega) c}{3} \bigl[ -\left(3 - 4\mathscr{P}_{00} \right) f_{\rm E, 2 m} \notag \\
    & ~~~+ \sqrt{3} \left(1 + f_{\text{I}, 0 0} \right) \mathscr{P}_{2 m} \bigr] \mbox{.}
\end{align}
\end{subequations}

\section{Solution for the brightness temperature}
\label{sec:results}

In this section, we collect the results of the previous sections, and derive their effect on observables i.e. the $21$-cm brightness temperature fluctuations.

Let us first consider the Boltzmann equation [Eq.~\eqref{eq:boltzmann}]. It is useful to define a few quantities to facilitate its solution and interpretation.

\begin{table}[t]
  \caption{\label{tab:sizes}Sizes of terms. They are classified as follows - \\
A: Terms included in the usual, lowest-order calculation. \\ B: Terms relevant to the effect under consideration. \\ C: Other terms of the same order.}
  \begin{ruledtabular}
    \begin{tabular}{l c c c c}
      Quantity & & & Sizes of relevant constituents & \\
      & & A & B & C \\
      \hline
      $f_{\text{I}, 0 0}(\mathcal{X})$ & & $T_\gamma/T_* + \left( \right) \tau$ & & $\left( \right) \tau^2$ \\
      $f_{\text{I}, 2 m}(\mathcal{X})$ & & $\left( \right) \delta \tau$ & $\left( \right) \delta \tau^2 $ & \\
      $f_{{\rm E}, 2 m}(\mathcal{X})$ & & & & $ \left( \right) \delta \tau^2 $ \\
      $\mathscr{P}_{2 m}$ & & & $\left( \right) \delta \tau $ &
    \end{tabular}
  \end{ruledtabular}
\end{table}

First, the optical depth $\tau$ of the neutral hydrogen gas is proportional to the absorption cross section integrated over the line. For a given Hubble rate $H$, and a peculiar velocity along the line of sight $v_{||}$,
\begin{align}
  \tau & = \frac{\pi^2 c^3 n_{\rm H} x_{1{\rm s}} A \left(3 - 4 \mathscr{P}_{0 0} \right)}{H \omega_{\rm hf}^3 [ 1 + (1/H) ({\rm d} v_{||}/{\rm d} r_{||}) ] } \nonumber \\
  & = 9.7 \times 10^{-3} \times x_{1{\rm s}} \left( \frac{T_\gamma}{T_{\rm s}} \right) \left[ 1 + \frac{4}{3} \delta \right] \left( \frac{\Omega_{\rm b} h^2}{0.022} \right)\notag \\
  & ~~~\times  \left( \frac{\Omega_{\rm m} h^2}{0.143} \right)^{-1/2} \left( \frac{1 - Y_{\rm He}}{0.75} \right) \left( \frac{1 + z}{10} \right)^{1/2} \mbox{.} \label{eq:opticaldepth}
\end{align}
This expression is correct to first order in the fluctuation $\delta$, and assumes that the slow variation of factors of $\omega$ in front of the absorption profile in Eq.~\eqref{eq:crosssection} can be neglected. Expression \eqref{eq:opticaldepth} is the optical depth for the monopole, since it is derived by averaging out the dependence of the velocity-gradient on direction.

Next is the cumulative function $\mathcal{X}(\omega)$ for the absorption profile $\phi(\omega)$, which is defined as
\begin{equation}
  \mathcal{X}(\omega) = \int_{-\infty}^{\omega} {\rm d} \omega^\prime \phi(\omega^\prime) \mbox{.}
\end{equation}
It is convenient to express the frequency dependence of quantities in terms of $\mathcal{X}$, which varies between $0$ and $1$ from the red- to the blue-side of the line. The boundary conditions for the moments are fixed on the blue side of the line i.e. at $\mathcal{X}=1$:
\begin{equation}
  \label{eq:boundaryconditions}
    f_{\text{I}, 0 0}  = f_\gamma \approx \frac{T_\gamma}{T_*} 
 {\rm~~and~~}
    f_{{\rm I}, 2 m}  =
    f_{{\rm E}, 2 m}  = 0 {\rm~~at~~}{\mathcal X}=1.
\end{equation}
Finally, the $21$-cm brightness temperature fluctuation relative to the CMB, $\delta T_{\rm b}$, is defined via the phase-space density on the red side of the line:
\begin{align}
  \delta T_{\rm b}(\hat{\bm n}) & = \frac{T_*}{1+z} \left( f_{\text{I}}(\mathcal{X}=0,\hat{\bm n}) - f_{\gamma} \right) \notag \\
  & \approx \frac{T_*}{1+z} \left( f_{\text{I}}(\mathcal{X}=0,\hat{\bm n}) - \frac{T_\gamma}{T_*} \right) \mbox{.} \label{eq:brightnessT}
\end{align}
Before we write down the form of the Boltzmann equation, it is worthwhile to note the sizes of various relevant terms. Table \ref{tab:sizes} shows the sizes of the relevant pieces, and summarizes the estimates made in Sec.~\ref{sec:estimate}.

We solve for the phase-space density in the steady state approximation. This holds if the time taken for the photon to redshift through the line is much smaller than a Hubble time, which is the case for a narrow line. Thus we safely neglect the time-derivatives in the free-streaming term [Eq.~\eqref{eq:photonfreestream}], and take the source terms from Eq.~\eqref{eq:photoncollision}. With the above definitions and assumptions, the Boltzmann equations for the various moments simplify to
\begin{subequations}
  \label{eq:boltzmannsteady}
  \begin{align}
    \frac{\partial f_{\text{I}, 0 0}}{\partial \mathcal{X}} & = \tau \left[ f_{\text{I}, 0 0} - \frac{T_{\rm s}}{T_*} \right] \mbox{,} \label{eq:f00boltzmann} \\
    \frac{\partial f_{\text{I}, 2 m}}{\partial \mathcal{X}} & = \tau \left[ f_{\text{I}, 2 m} - \frac{2 \sqrt{2}}{3} \frac{T_\gamma T_{\rm s}}{T_*^2} \mathscr{P}_{2 m} \right] \notag \\
    & ~~~+ \frac{2}{3} \delta \frac{\partial f_{\text{I}, 0 0}}{\partial \mathcal{X}} \sqrt{\frac{4\pi}{5}} Y_{2 m}(\hat{\bm k}) \mbox{,~and} \label{eq:f2mboltzmann} \\
    \frac{\partial f_{{\rm E}, 2 m} }{\partial \mathcal{X}} & = \tau \left[ f_{{\rm E}, 2 m} - \frac{4\sqrt{3}}{3} \, \frac{T_\gamma T_{\rm s}}{T_*^2} \mathscr{P}_{2 m} \right] \mbox{.} \label{feboltzmann}
  \end{align}
\end{subequations}
The velocity-gradient contribution to the optical depths of the quadrupoles is different, but these moments vanish in the absence of fluctuations. Hence Eq.~\eqref{eq:boltzmannsteady} is correct to first order in the overdensity $\delta$. The simplifications here use the sizes of various terms from Table \ref{tab:sizes}, the relation of Eq.~\eqref{eq:spintemperature} between the excited state occupancy $\mathscr{P}_{0 0}$ and the spin temperature $T_{\rm s}$, and neglect spontaneous emission contributions.

The Boltzmann equation must be solved along with the evolution equations for the density matrix of the hydrogen atoms. We obtain these from Secs.~\ref{sec:dmevolnradiative} and \ref{sec:otherproc}, and include the effects of interaction with radio photons, [Sec.~\ref{sec:dmevolnradiative}], optical pumping by Lyman-$\alpha$ photons [Sec.~\ref{subsec:opticalpumping}], collisions with other hydrogen atoms [Sec.~\ref{subsec:collisions}] and precession within an external magnetic field [Sec.~\ref{subsec:magfield}]. Similar to the phase-space density, we solve for the various parts of the density matrix under the steady state approximation.

First, we obtain the evolution of the excited state occupancy $\mathscr{P}_{0 0}$ (or alternatively, the spin temperature $T_{\rm s}$) by summing Eqs.~\eqref{eq:p00evolrad}, \eqref{eq:p00evolcoll} and \eqref{eq:p00dotlya} and equating the result to zero:
\begin{align}
  \dot{\mathscr{P}}_{0 0} & = A \left[ - \mathscr{P}_{00} + \left(3 - 4 \mathscr{P}_{00} \right) \overline{f_{\text{I}, 0 0}} \right] \notag \\
  & ~~~- \frac{32\pi \lambda_{\text{Ly}\alpha}^2 \gamma_{2{\rm p}}}{9} \tilde{S}_\alpha J_\alpha \left( \mathscr{P}_{00} - \frac{3}{4} + \frac{3}{16} \frac{T_*}{T_{\rm c, eff}} \right) \notag \\
  & ~~~- 4\kappa(1 \mhyphen 0) n_{\rm H} \left( \mathscr{P}_{00} - \frac{3}{4} + \frac{3}{16}\frac{T_*}{T_{\rm k}} \right) = 0 \mbox{.} \label{eq:p00dot}
\end{align}
In a similar manner, we obtain the equation for the evolution of the alignment tensor $\mathscr{P}_{2 m}$ by summing Eqs.~\eqref{eq:p2mevolrad}, \eqref{eq:p2mevolcoll}, \eqref{eq:p2mdotlya} and \eqref{eq:precession}. It is most convenient to continue in the coordinate system used in Sec.~\ref{subsec:magfield}, with the $z-$axis along the direction of the magnetic field; In this system, the angular indices $j m$ are not mixed:
\begin{align}
\!\!\!\!
  \dot{\mathscr{P}}_{2 m} & = A \Bigl[ - \frac{T_\gamma}{T_*} \mathscr{P}_{2 m} + \frac{3}{20 \sqrt{2}} \frac{T_*}{T_{\rm s}} \overline{f_{\text{I}, 2 m}} \Bigr] \notag \\
  & ~~~- 3.607 \pi \lambda_{\text{Ly}\alpha}^2 \gamma_{2{\rm p}} \tilde{S}_{\alpha, (2)} J_\alpha \mathscr{P}_{2 m}  \notag \\
  & ~~~
  - n_{\rm H} \kappa^{(2)}(1 \mhyphen 0) \mathscr{P}_{2 m}
  + i \frac{m}{2} \frac{g_{\rm e} \mu_{\rm B}}{\hbar} B \mathscr{P}_{2 m} \approx 0 \mbox{.} \label{eq:p2mdot}
\end{align}
As earlier, the above equation neglects spontaneous emission and is correct up to the sizes of terms from Table \ref{tab:sizes}. We carry out the averages over the line-profile in Eqs.~\eqref{eq:p00dot} and \eqref{eq:p2mdot} using
\begin{equation}
  \overline{f} = \int_{-\infty}^{\infty} {\rm d} \omega f(\omega) = \int_{0}^{1} {\rm d} \mathcal{X} f(\mathcal{X}) \mbox{.}
\end{equation}
Equations \eqref{eq:p00dot} and \eqref{eq:f00boltzmann} together determine the spin temperature $T_{\rm s}$ and the intensity monopole $f_{\text{I}, 0 0}$, which is given in terms of the former by
\begin{equation}
  f_{\text{I}, 0 0}(\mathcal{X}) 
  = 
  \frac{1}{T_*} \left[ T_{\rm s} +  \left( T_\gamma - T_{\rm s} \right) e^{-\tau(1 - \mathcal{X})} \right] \mbox{.} \label{eq:monopole}
\end{equation}
Likewise, we use Eqs.~\eqref{eq:p2mdot} and \eqref{eq:f2mboltzmann} to solve for the alignment tensor $\mathscr{P}_{2 m}$, and the intensity quadrupole $f_{\text{I}, 2 m}(\mathcal{X})$ in a simultaneous manner. They are given by the following solutions, which are correct to the orders in Table \ref{tab:sizes}:
\begin{align}
  \mathscr{P}_{2 m}
  & = \frac{1}{20 \sqrt{2}} \frac{T_*}{T_\gamma} \left( 1 - \frac{T_\gamma}{T_{\rm s}} \right) \frac{ \tau }{1 + x_{ \alpha, (2) } + x_{{\rm c}, (2)} - i m x_{\rm B}} \notag \\
  & ~~~\times \delta \sqrt{\frac{4\pi}{5}} Y_{2 m}(\hat{\bm k})
\end{align}
and
\begin{align}
  f_{\text{I}, 2 m}(\mathcal{X})
  & = \frac{T_{\rm s}}{T_*} \left( 1 - \frac{T_\gamma}{T_{\rm s}} \right) \biggl[ \frac{1}{30} \frac{ \tau }{1 + x_{\alpha, (2)} + x_{{\rm c}, (2)} - i m x_{\rm B}} \notag \\
    & ~~~+ \frac{2}{3} \left( 1 - \tau(1 -  \mathcal{X}) \right) \biggr] \delta \tau (1 - \mathcal{X}) \sqrt{\frac{4\pi}{5}} Y_{2 m}(\hat{\bm k}) \mbox{,} \label{eq:quadrupole}
\end{align}
where the quantities $x_{\alpha, (2)}$, $x_{{\rm c}, (2)}$ and $x_{\rm B}$ parametrize the rates of depolarization by optical pumping and collisions, and precession relative to radiative depolarization. They are given by
\begin{align}
  x_{\alpha, (2)} & = \frac{3.607 \pi \lambda_{\text{Ly}\alpha}^2 \gamma_{2{\rm p}} T_*}{A T_\gamma} \tilde{S}_{\alpha, (2)} J_\alpha \notag \\
  & = 0.073 \tilde{S}_{\alpha, (2)} \left( \frac{1+z}{10} \right)^{-1} 
  \notag \\ & ~~\times
  \left( \frac{J_\alpha}{10^{-12} \textrm{cm}^{-2} \textrm{sr}^{-1} \textrm{s}^{-1} \textrm{Hz}^{-1}} \right) \mbox{,} \label{eq:xalpha} \\
  x_{\rm c, (2)} & = \kappa^{(2)}(1 \mhyphen 0) \frac{n_{\rm H} T_*}{A T_\gamma} \notag \\
  & = 2 \times 10^{-3} \left( \frac{1+z}{10} \right)^2 \left( \frac{\kappa^{(2)}(1 \mhyphen 0)}{1.3 \times 10^{-11} \textrm{cm}^{3} \textrm{s}^{-1} } \right) \mbox{,~and} \label{eq:xc} \\
  x_{\rm B} & = \frac{g_{\rm e} \mu_{\rm B} T_*}{2 \hbar A T_\gamma} B \notag \\
  & = 0.698 \left( \frac{1+z}{10} \right)^{-1} \left( \frac{B}{10^{-19} \textrm{G}} \right) \mbox{.} \label{eq:xb}
\end{align}
In the equation above, $B$ is the local value of the magnetic field.

We compute the brightness temperature fluctuation, $\delta T_{\rm b}$, from Eq.~\eqref{eq:brightnessT}, wherein the phase-space density is given by the sum of the monopole and quadrupole from Eqs.~\eqref{eq:monopole} and \eqref{eq:quadrupole} respectively. We get the following expression, which is one of the main results of this paper:
\begin{align}
  \delta T_{\rm b}(\hat{\bm n}) &= \left( 1 - \frac{T_\gamma}{T_{\rm s}} \right) x_{1{\rm s}} \left( \frac{1+z}{10} \right)^{1/2} \notag \\ 
  & \hspace{10pt}\times \biggl[ 26.4 \ {\rm mK} \Bigl\{ 1 + \left(1 + (\hat{\bm k} \cdot \hat{\bm n})^2 \right)\delta \Bigr\} \notag \\
  & \hspace{20pt}- 0.128 \ {\rm mK} \left( \frac{T_\gamma}{T_{\rm s}} \right) x_{1{\rm s}} \left( \frac{1+z}{10} \right)^{1/2} \notag \\
  & \hspace{20pt}\times \Bigl\{ 1 + 2 \left(1 + (\hat{\bm k} \cdot \hat{\bm n})^2 \right)\delta \notag \\
  & \hspace{30pt}- \frac{ \delta }{15} \sum_m \frac{4\pi}{5} \frac{Y_{2 m}(\hat{\bm k}) \left[ Y_{2 m} (\hat{\bm n}) \right]^* }{1 + x_{ \alpha, (2) } + x_{{\rm c}, (2)} - i m x_{\rm B}} \Bigr\} \biggr] \mbox{.} \label{eq:tbsoln}
\end{align}
Equation ~\eqref{eq:xb} offers a rough guide to estimate the strengths of magnetic fields to which the method outlined in this paper is most sensitive. We must keep in mind that the coefficient $x_{\rm B}$ only measures the strength of the precession relative to radiative depolarization, and a full analysis of the discriminating power of this method must estimate the sizes of Ly$\alpha$ and collisional depolarization, or the coefficients $x_{\alpha, (2)}$ and $x_{\rm c, (2)}$ in Eqs.~\eqref{eq:xalpha} and \eqref{eq:xc}. The second paper in this series studies this in more detail. For now, we note that field strengths of ${\cal O}(10^{-19} \ {\rm G})$ at redshifts of $z \sim 10$ are associated with $x_{\rm B} \sim 1$.

Given this scale of field strengths, we identify two physical regimes -- one with weaker fields, and one with much stronger ones. We use the weak-field limit of Eq.~\eqref{eq:tbsoln} to make contact with the intuitive picture laid out in Sec.~\ref{sec:estimate}. If we consider a magnetic field that is coherent on larger scales than the 21--cm fluctuations of interest and thus effectively homogenous, and take the limit of $x_{\rm B} \rightarrow 0$ in Eq.~\eqref{eq:tbsoln}, we obtain the following coordinate independent response to the weak field:
\begin{align}
  \frac{ {\rm d} \delta T_{\rm b} }{ {\rm d} B } (\hat{\bm n}) & = 1.786 \times 10^{17} \ \frac{\rm mK}{\rm G} ~ [ \hat{\bm B} \cdot ( \hat{\bm k} \bm \times \hat{\bm n} ) ] ( \hat{\bm n} \cdot \hat{\bm k} ) \nonumber \\
  & ~~~\times \left( 1 - \frac{T_\gamma}{T_{\rm s}} \right) x_{1{\rm s}}^2 \left( \frac{T_\gamma}{T_{\rm s}} \right) \frac{ \delta }{ (1 + x_{\alpha, (2)} + x_{{\rm c}, (2)})^2 } \mbox{.} \label{eq:linearB}
\end{align}
In the geometry of Fig.~\ref{fig:schematic}, the direction to the observer is $\hat{\bm n} = -\hat{\bm y}$. If we substitute this in the above equation, we recover the angular structure of the correction to the brightness temperature in Sec.~\ref{sec:estimate}, in particular, the form of Eq.~\eqref{eq:estimate}. The latter only accounted for the radiative decay of the magnetic moment, while Eq.~\eqref{eq:linearB} includes the additional effect of collisions and optical pumping through the dimensionless factors of $x_{\alpha, (2)}$ and $x_{ {\rm c}, (2)}$.

We realize the complementary strong field limit by taking the limit $x_{\rm B} \rightarrow \infty$ in Eq.~\eqref{eq:tbsoln}. The change in brightness temperature over the case with no external magnetic field is
\begin{align}
  ~ ~ ~ & \!\!\!\!
  \delta T_{\rm b}( \hat{\bm n} ) \vert_{x_{\rm B} \rightarrow \infty} - \delta T_{\rm b}( \hat{\bm n} ) \vert_{x_{\rm B} = 0 } \nonumber \\
  & = 8.53 \ \mu{\rm K} \times P_{2}( \hat{\bm k} \cdot \hat{\bm B} ) P_{2}( \hat{\bm n} \cdot \hat{\bm B} ) \nonumber \\
  & ~~~\times \left( 1 - \frac{T_\gamma}{T_{\rm s}} \right) x_{1{\rm s}}^2 \left( \frac{1+z}{10} \right) \left( \frac{T_\gamma}{T_{\rm s}} \right) \frac{ \delta }{1 + x_{ \alpha, (2) } + x_{{\rm c}, (2)} } \mbox{.} \label{eq:largeB}
\end{align}
From the above expression, we see that the effect saturates at large values of the magnetic field strength. However, we observe that it is still possible to reconstruct the direction of the magnetic field in the plane of the sky using the form of the isotropy breaking in $\hat{\bm k}$ space. The correction is roughly three orders of magnitude fainter than the raw 21-cm brightness even for the optimal range of $\hat{\bm k}$, $\hat{\bm B}$, and $J_\alpha$. However, it should be noted that it is exactly in phase with the conventional brightness temperature fluctuations---that is, it traces the same underlying density field $\delta$ and is changing the coefficient in front of this. Thus its effect on the power spectrum is of order $10^{-3}$, not $10^{-6}$ (as would be the case if the magnetic field correction were a new random field, independent of the density but with an amplitude three orders of magnitude smaller).

\section{Summary and Conclusions}
\label{sec:summary}
 
In this study, we propose a new method to probe magnetic fields present in the universe prior to and during the early stages of cosmic reionization. The method relies on the spin-polarization of the triplet state of the hyperfine sublevels of neutral hydrogen by an anisotropic radiation field near the energy of the $21$-cm transition. These anisotropies naturally arise in the early universe due to density fluctuations in the high redshift gas. In the presence of an external magnetic field, the precession of these spin-polarized atoms changes the angular distribution of the emitted $21$-cm radiation at second order in optical depth. If the external magnetic fields are coherent over much larger length-scales than the 21-cm fluctuations, they are effectively homogenous and break the isotropy of the signal along the line of sight---this produces a characteristic signature in the two-point correlation function of the brightness-temperature fluctuations.

Due to the long lifetimes of the excited states of the hyperfine transition, this method is naturally optimal for measuring very weak magnetic fields ($\lesssim 10^{-19}$ G at a reference redshift $z\sim$20, or $\lesssim 10^{-21}$ G in comoving units). It thus raises the exciting possibility of probing seed fields that may have given rise to the magnetic fields observed in the present-day universe. As the background magnetic field increases, the effect saturates; however, even in the saturated case, it is possible to recover the field's direction, and a lower limit on its strength, as discussed in detail in Paper II.

In order to evaluate this effect, we present a detailed calculation of the coupled evolution of atomic and photon density matrices. We account for all the processes which affect the atomic magnetic moments, such as the Wouthuysen-Field effect, atomic collisions, and radiative decay. The main results are Eq.~\eqref{eq:tbsoln}, which includes the corrections to the brightness temperature due to all these effects, and Eqs.~\eqref{eq:linearB} and \eqref{eq:largeB}, which show the weak- and strong-field limits, respectively. This calculation provides a complete theoretical basis for understanding the microphysics of the hyperfine transition in the presence of external magnetic fields, and for calculating the effect of magnetic precession on 21-cm brightness-temperature signal.

The method we proposed here adds to the already exciting opportunities for the use of the 21-cm line as a probe of the early universe, and is in principle sensitive to extremely weak magnetic fields which are far beyond the reach of any other method (including other techniques based on the 21-cm radiation).
Paper II of this series \cite{2016arXiv160406327G} presents a formalism to search for evidence of magnetic fields in data from future 21-cm tomography surveys, both using the method presented here, and an extension that is adopted for fields that vary on the survey-scales. In Paper II, we also forecast sensitivity of future surveys to detecting any particular model for magnetic fields (regardless of their origin), and find that an array of dipole antennas, with a collecting area slightly larger than a square kilometer, is able reach $1\sigma$ sensitivity to detecting magnetic fields near saturation ($\sim10^{-21}$ comoving G at $z\sim$21), with about three years of integration time.
 
\begin{acknowledgments}
We would like to thank Peter Goldreich and Takeshi Kobayashi for some helpful conversations during the early part of this work. T.V. acknowledges support from the Schmidt Fellowship and the Fund for Memberships in Natural Sciences at the Institute for Advanced Study. V.G. gratefully acknowledges the support of the Friends of the Institute for Advanced Study in Princeton. During the duration of this work, T.V. and A.O. were supported by the International Fulbright Science and Technology Award, and C.H., A.M., A.O., and T.V. were supported by the David and Lucile Packard Foundation, the Simons Foundation, and the U.S. Department of Energy. C.H. is also supported by NASA.
\end{acknowledgments}

\appendix

\section{Conventions for spherical tensors}
\label{sec:conventions}

In this section, we lay out the conventions for spherical tensors we use in the body of the paper, and our reasons for adopting the same.

Consider a passive rotation around the $z$-axis by an angle $\alpha$, which connects two coordinate systems $S$ and $S^\prime$ as follows:
\begin{align}
  ( \theta, \phi ) \vert_{S} & = ( \theta, \phi - \alpha ) \vert_{S^\prime} \mbox{,}
\end{align}
where both sides refer to the same point on the unit sphere. Within quantum mechanics, the coefficients of a state and expectation values of spherical tensors transform with opposite signs:
\begin{align}
  c_m \vert_{S^\prime} & = e^{i m\alpha} c_m \vert_S \ {\rm ~with~ } \ \vert \psi \rangle = \sum_m c_m \vert m \rangle  \label{eq:statetrans}
  \end{align}
for states and
  \begin{align}
    \langle T^{(k)}_m \rangle \vert_{S^\prime} & = e^{-i m \alpha} \langle T^{(k)}_m \rangle \vert_{S}
     \label{eq:operatortrans}
\end{align}
for spherical tensors.
The spherical tensors of interest are the irreducible components of the matter density matrix ($\mathscr{P}_{j m}$), and the moments of the phase-space density matrix of the radiation [$(f_{\alpha \beta})_{j m}$]. They are defined in Eqs.~\eqref{eq:polmom} and \eqref{eq:momentexpansion}; these definitions transform in the manner of Eq.~\eqref{eq:operatortrans}.

Note that the definition of the multipoles of the radiation in Eq.~\eqref{eq:momentexpansion} differs from the usual convention adopted in cosmology literature, which omits the complex conjugate on its RHS. The latter considers these moments as state-coefficients rather than expectation values of spherical tensors. Considering that the majority of the calculations in this paper have an atomic physics flavor, our definition is convenient, though unconventional.

\section{Spherical Wave Basis for the Radiation's Phase-space Density Matrix}
\label{sec:sphwave}

The standard choice of basis for the EM field's expansion is one consisting of plane waves, whose defining characteristic is that they are eigenfunctions of the linear momentum and helicity of the EM field. This is the basis used in Sec.~\ref{subsec:photondm}. However, it is also possible to use eigenstates of the total angular momentum, parity and energy of the EM field as basis elements. This section expands on this, and details how to transform between these two bases.

Eigenstates of total angular momentum have the usual indices $j$ and $m$. They are classified as electric and magnetic type states depending on how they behave under a parity transformation---electric type states pick up a factor of $(-1)^j$, while those of the magnetic type pick up $(-1)^{j+1}$. The explicit form of these eigenstates is \cite{Lifshitz71}
\begin{align}
\!\!\!\!\!  \bm{A}_{\omega,j m}^{(\lambda)}({\bm r}) & = \int \frac{{\rm d}^3 {\bm k}_\gamma}{(2\pi)^3} \bm{A}_{\omega,jm}^{(\lambda)}({\bm k}_\gamma) e^{i{\bm k}_\gamma \cdot{\bm r}}, \hspace{10pt} \lambda = {\rm E,M} \\
\!\!\!\!\!  \bm{A}_{\omega,j m}^{(\lambda)}(\bm{k}_\gamma) & = 4\pi^2 \left( \frac{\hbar c^3}{\omega^3} \right)^{1/2} \!\!\delta(k_\gamma - \omega/c) {\bm Y}_{jm}^{(\lambda)} (\hat{\bm n}) \mbox{,~\!and}\! \label{eq:photonwavefunction} \\
\!\!\!\!\!  {\bm Y}_{j m}^{(\lambda)}(\hat{\bm n}) & = \begin{cases} \frac{1}{\sqrt{j(j+1)}} \nabla_{\hat{\bm n}} Y_{jm} & \lambda = {\rm E} \\
    \frac{1}{\sqrt{j(j+1)}} \hat{\bm n} \times \nabla_{\hat{\bm n}} Y_{jm} & \lambda = {\rm M}
  \end{cases} \mbox{,}
\end{align}
where $\hat{\bm n} = \hat{\bm k}_\gamma$ is the direction of propagation and the index $j$ runs over integers greater than zero, while $m$ runs over integers from $-j$ to $j$.

We expand the vector potential $\bm A$ in the same manner as in Eq.~\eqref{eq:aexpand}.
\begin{align}
  \bm{A}(\bm{r}) & = \sum_{j,m} \int  \bigl[ \bigl\{ a_{j m}^{({\rm E})}(\omega) \bm{A}_{\omega,j m}^{({\rm E})}(\bm r) + a_{j m}^{({\rm M})}(\omega) \bm{A}_{\omega,j m}^{({\rm M})}(\bm{r}) \bigr\} \notag \\
  & ~~~+ {\rm h.c.} \bigr]\,{\rm d} \omega \mbox{,} \label{eq:asphwaveexpand}
\end{align}
where the operators $a_{\omega,j m}^{(e/m)}$ and $a_{\omega,j m}^{(e/m)}{^\dagger}$ are annihilation and creation operators for photons of the electric and magnetic type. Operators for photons of the same type have the following commutation relations:
\begin{align}
  [a_{j m}(\omega),a_{j^\prime m^\prime}^\dagger(\omega^\prime)] & = \delta(\omega-\omega^\prime) \delta_{j,j^\prime} \delta_{m,m^\prime} \notag{\rm ~and}  \\
  [a_{j m}(\omega),a_{j^\prime m^\prime}(\omega^\prime)] & = [a_{j m}^{\dagger}(\omega),a_{j^\prime m^\prime}^\dagger(\omega^\prime)] = 0 \mbox{,}\label{eq:commutation}
\end{align}
while those of different types commute with each other.

The phase-space density matrix in this basis can be defined in the same manner as in Eq.~\eqref{eq:falphabeta} for the plane wave basis:
\begin{align}
 \langle a_{j m}^{(\lambda)\dagger}(\omega) \ a_{j^\prime m^\prime}^{(\lambda^\prime)}(\omega^\prime) \rangle & = f_{m^\prime, m}^{(\lambda^\prime j^\prime)(\lambda j)}(\omega) \ \delta(\omega - \omega^\prime) \label{eq:rhojmjm}
\end{align}
for $\lambda,\lambda^\prime = {\rm E,M}$.

At this stage, it is worthwhile to examine the general considerations leading to the forms of the density matrices in the two bases. Phase coherence between frequencies separated by $\Delta \omega$ leads to oscillatory features on time-scales of $\Delta t \sim 1/\Delta \omega$. If the time-interval $\Delta t$ over which the statistical properties of the radiation field are stationary is sufficiently long, the width of the two-point function in frequency space is $\sim 1/\Delta t \rightarrow 0$. Thus the $\delta$-function in the definition in the spherical wave basis [Eq.~\eqref{eq:rhojmjm}] is a consequence of time-translation invariance. 

The $\delta$-function in the definition in the plane wave basis [Eq.~\eqref{eq:falphabeta}] is a consequence of invariance under spatial translations, the argument paralleling the one for time-translation invariance above. It is relatively simple to express a state given in the plane wave basis in the spherical one, but the inverse transformation involves averaging over the positions of the interacting atoms to recover translational invariance. This is dealt with in greater detail in Sec.~\ref{subsec:collisionterm}.

In the rest of this section, we describe the transformation from the plane wave basis (the $f_{X,j m}$s) to the spherical wave one (the $f_{m,m^\prime}^{(\lambda j)(\lambda^\prime j^\prime)}$s) centered at the position of a hydrogen atom interacting with the radiation. The transformation is
\begin{align}
  f_{m,m^\prime}^{(\lambda j)(\lambda^\prime j^\prime)}(\omega) & = \sum_{\alpha,\beta} \int {\rm d}^2 \bm n \, f_{\alpha \beta}(\omega, \hat{\bm n}) \notag \\
  & ~~~\times 
  \bigl[ \bm{e}_{(\alpha)} \cdot {\bm Y}_{jm}^{(\lambda)}{^*} \bigr] (\hat{\bm n})
  \bigl[ \bm{e}_{(\beta)}^{*} \cdot {\bm Y}_{j^\prime m^\prime}^{(\lambda^\prime)} \bigr] (\hat{\bm n}) \mbox{.} \label{eq:pltosphtransform}
\end{align}
The normalization is such that if the radiation is unpolarized and isotropic (e.g. a thermal state), the elements of the phase-space density matrix are
\begin{equation}
  f_{m, m^\prime}^{(\lambda j)(\lambda^\prime j^\prime)}(\omega) = \begin{cases} f_{\text{I}, 0 0}(\omega) \, \delta_{j,j^\prime} \delta_{m,m^\prime} & \text{ if } \lambda = \lambda^\prime \\
    0 & \text{ if } \lambda \ne \lambda^\prime
  \end{cases} \ \mbox{.}
\end{equation}
We further simplify the angular integral in the transformation of Eq.~\eqref{eq:pltosphtransform} using the moments of the phase-space density matrix in the plane wave basis [Eq.~\eqref{eq:momentexpansion}], and the Clebsch-Gordan rule for evaluating the angular integral of the product of three spherical harmonics \cite{Varshalovich88}. 

The M1--M1 block of the phase-space density matrix contributes to the evolution of the atom density matrix [see Sec.~\ref{sec:dmevolnradiative}]. We derive its explicit form for arbitrarily polarized radiation by simplifying Eq.~\eqref{eq:pltosphtransform}:
 \begin{align}
 f_{m, m^\prime}^{({\rm M}1)({\rm M}1)}(\omega) & = \frac{3}{2} \sum_{j,m_2} \sum_{\alpha,\beta} \, \alpha \beta (-1)^{\alpha - m^\prime} (f_{\alpha \beta})_{j m_2}(\omega) \notag \\
 & ~~~\times \tj{1}{1}{j}{-\alpha}{\beta}{\alpha-\beta}
 \tj{1}{1}{j}{-m}{m^\prime}{-m_2}  \mbox{.} \label{eq:m1m1block}
 \end{align}
This $3 \times 3$ block is equivalently described in terms of its irreducible components $\mathcal{F}_{j m}(\omega)$  of ranks $j = \{0,1,2\}$, in exactly the same manner as the matter density matrix $\rho_{m_1 m_2}$ in Eqs.~\eqref{eq:polmom} and \eqref{eq:polmom2}:
\begin{align}
\mathcal{F}_{j m}(\omega) & = \sqrt{(2j+1)3} \sum_{m_1,m_2} (-1)^{1-m_2} \tj{1}{j}{1}{-m_2}{\mu}{m_1}  \notag \\
& ~~~\times f_{m_1, m_2}^{({\rm M}1)({\rm M}1)}(\omega) \mbox{,} \label{eq:fjm}
\end{align}
with the inverse relation
\begin{align}
f_{m_1 m_2}^{({\rm M}1)({\rm M}1)}(\omega) & = \sum_{j m} \sqrt{\frac{2j+1}{3}} (-1)^{1-m_2}\tj{1}{j}{1}{-m_2}{m}{m_1} \notag \\
& ~~~\times \mathcal{F}_{j m}(\omega) \mbox{.} \label{eq:fjminv}
\end{align}
Substitution in Eq.~\eqref{eq:m1m1block} gives the explicit forms of these irreducible components
\begin{subequations}
\label{eq:photonpolmoments}
\begin{align}
  \mathcal{F}_{0 0}(\omega) & = 3 f_{\text{I}, 0 0}(\omega) \mbox{,} \\
  \mathcal{F}_{1 m}(\omega) & = \sqrt{\frac{3}{2}} f_{{\rm V}, 1 m}(\omega) \mbox{,~and} \\
  \mathcal{F}_{2 m}(\omega) & = \frac{3}{5\sqrt{2}} \left[ f_{\text{I}, 2 m}(\omega) + \sqrt{6} \, f_{{\rm E}, 2 m}(\omega) \right] \mbox{.}
\end{align}
\end{subequations}

\section{Three-point functions of the atoms and the radiation field}
\label{sec:3ptfn}

Three-point functions of the atom and the radiation field affect the evolution of the atoms' density matrix $\rho$ and the radiation's phase-space density matrix $f$. In this section, we derive expressions for their contribution.

The unperturbed Hamiltonians for the hydrogen atoms and radiation are
\begin{align}
  H_{\rm hf} & = E_0 \vert 0 0 \rangle \langle 0 0 \vert + E_1 \sum_m \vert 1 m \rangle \langle 1 m \vert \mbox{,} \label{eq:atomhamilt} \\
  H_{\gamma} & = \sum_{j, m, \lambda} \int {\rm d} \omega \, \hbar \omega \, a_{j m}^{(\lambda)}{^\dagger}(\omega) a_{j m}^{(\lambda)}(\omega) \mbox{,} \label{eq:photonhamilt}
\end{align}
where $E_0$ and $E_1$ are the energies of the singlet and triplet levels. The zero-point energy has been left out of Eq.~\eqref{eq:photonhamilt}.

A three-point function is the expectation value of an operator consisting of the product of creation and annihilation operators for the hydrogen atoms and for the radiation. This function's evolution is governed by the operator's commutator with the total Hamiltonian:
\begin{align}
  ~~ & \!\!\!
  \frac{\rm d}{{\rm d} t} \left\langle \vert 1 m_1 \rangle \langle 0 0 \vert a_{j m}^{(\lambda)}(\omega) \right\rangle \notag \\
  & = i ( \omega_{\rm hf} - \omega ) \left\langle \vert 1 m_1 \rangle \langle 0 0 | a_{j m}^{(\lambda)}(\omega) \right\rangle \notag \\
  & ~~~+ \frac{i}{\hbar} \left\langle \bigl[ H_{\rm{hf},\gamma}, \vert 1 m_1 \rangle \langle 0 0 \vert a_{j m}^{(\lambda)}(\omega) \bigr] \right\rangle \mbox{.} \label{eq:threepointevolution}
\end{align}
Assuming that the interaction is turned on at $t=0$, the formal solution to Eq.~\eqref{eq:threepointevolution} is
\begin{equation}
\begin{split}
  & \left\langle \vert 1 m_1 \rangle \langle 0 0 \vert a_{j m}^{(\lambda)}(\omega) \right\rangle = \mathcal{C} e^{i(\omega_{\rm hf} - \omega)t} \\
  & ~~~+ \frac{i}{\hbar} \int_0^t {\rm d} t^\prime e^{-i(\omega_{\rm hf} - \omega)(t^\prime - t)} \left\langle \left[ H_{\rm{hf},\gamma}, \vert 1 m_1 \rangle \langle 0 0 \vert a_{j m}^{(\lambda)}(\omega) \right] \right\rangle \mbox{.}
  \end{split}
\end{equation}
If the expectation value in the integrand of the second term varies slowly with time, the exponential dominates the integral and results in a $\delta$-function which picks out the frequency resonant with the level gap. This behaves like a rate term when the three-point function is input to an evolution equation (the Fermi golden rule). The first term does not lead to such a secular rate contribution. We have the identity
\begin{align}
  ~~ & \!\!\!
  \left\langle \vert 1 m_1 \rangle \langle 0 0 \vert a_{j m}^{(\lambda)}(\omega) \right\rangle \notag \\
  & = \frac{i}{\hbar} \pi \delta(\omega - \omega_{\rm hf}) \left\langle \left[ H_{\rm{hf},\gamma}, \vert 1 m_1 \rangle \langle 0 0 \vert a_{j m}^{(\lambda)}(\omega) \right] \right\rangle \mbox{.} \label{eq:3ptfna}
\end{align}
We use the form of the interaction Hamiltonian from Eq.~\eqref{eq:M1Hamilt} to evaluate the last expectation value. Expanding the commutator leads to 4-point functions, which we separate into atom and photon density matrices under the assumption of weak interactions:
\begin{align}
  ~~ & \!\!\!
  \left\langle \left[ H_{\rm{hf},\gamma}, \vert 1 m_1 \rangle \langle 0 0 \vert a_{j m}^{(\lambda)}(\omega) \right] \right\rangle \notag \\
  & = \sum_{m_2, m^\prime} V_{m_2 a, m^\prime}^*(\omega) \Bigl[ \delta_{m_1 m_2} \, \rho_{a a} \, f_{m, m^\prime}^{(\lambda j),({\rm M} 1)}(\omega) \notag \\
   & ~~~- \rho_{m_2 m_1} \Bigl\{ \delta_{(\lambda)({\rm M})} \delta_{j 1} \delta_{m m^\prime} + f_{m, m^\prime}^{(\lambda j)({\rm M} 1)}(\omega) \Bigr\} \Bigr] \mbox{.}
\end{align}

\bibliography{references}

\end{document}